\documentclass[11pt]{article}
\usepackage{graphicx} 
\usepackage{amsmath}
\usepackage{amssymb}
\usepackage{booktabs}
\usepackage{geometry}
\usepackage{multirow}
\usepackage{subcaption}
\usepackage{float}
\usepackage[colorlinks=true, linkcolor=blue, citecolor=blue, urlcolor=blue]{hyperref}
\usepackage{tikz}
\usetikzlibrary{arrows.meta,decorations.pathreplacing,backgrounds}
\definecolor{rxcolor}{RGB}{42,100,150}
\definecolor{rzcolor}{RGB}{177,112,36}
\definecolor{xxcolor}{RGB}{112,77,145}
\definecolor{wirecolor}{RGB}{55,62,70}
\definecolor{panelcolor}{RGB}{246,248,251}
\usepackage{siunitx}
\usepackage{etoolbox}
\robustify\bfseries
\newcolumntype{M}{S[table-format=1.4]@{\,$\pm$\,}S[table-format=1.4]}
\title{QPAT: A Universal Quantum Machine Learning Framework for Quantum-Assisted Training and Classical Inference}
\author{Yixiong Chen}
\date{}

\begin{document}

\maketitle

\begin{abstract}
Deep learning models often struggle with sharp local optima and overfitting in limited-data regimes, and exhibit limited capacity for capturing high-order, complex probability distributions. Quantum Machine Learning (QML) models, by operating in exponentially large Hilbert spaces, can efficiently represent such distributions and provide favorable inductive biases for learning, offering a promising pathway to overcome these limitations. However, their practical deployment in the Noisy Intermediate-Scale Quantum (NISQ) era is severely hindered by gradient coupling, barren plateaus, data-loading bottlenecks, and high inference latency. To bridge classical AI and QML, we propose QPAT (QCBM-Prior-Assisted Training), a universal framework that employs a Quantum Circuit Born Machine (QCBM) as a learnable generative prior to provide a task-aligned structured inductive bias during classical model training. The framework can be seamlessly integrated into any classical backbone, including LSTMs, CNNs, and Transformers. We introduce a decoupled fitness-based scoring mechanism and a stochastic ``Drop-Prior'' mechanism to implicitly distill quantum structural knowledge into the classical weights during training. Evaluated extensively on six benchmarks spanning time-series forecasting, computer vision, and natural language processing, our framework consistently outperforms purely classical backbones in both predictive performance and stability across random seeds; ablation studies further show that these gains are not reproduced by unstructured random priors or a classically learned generative prior. Remarkably, our approach enables completely classical inference, bypassing quantum latency while preserving quantum-enhanced accuracy, paving a highly practical path for QML deployment.
\end{abstract}

\section{Introduction}
Deep learning has enabled substantial advances in representation learning for visual and sequential data~\cite{lecun2015deep, he2016resnet, hochreiter1997lstm, vaswani2017attention}. Nevertheless, classical models continue to face challenges in optimization and generalization, including overfitting and unreliable generalization in limited-data regimes~\cite{srivastava2014dropout}, as well as a limited capacity for modeling high-order, complex probability distributions. These challenges have motivated growing interest in Quantum Machine Learning (QML) as an alternative route for representation learning and probabilistic modeling~\cite{biamonte2017quantum,cerezo2022challenges}. Quantum feature-space methods embed data into exponentially large Hilbert spaces, where complex patterns can become linearly separable, providing favorable inductive biases for learning from limited data~\cite{schuld2019quantum,havlicek2019supervised}, while parameterized quantum circuits provide trainable models for both discriminative and generative tasks~\cite{benedetti2019parameterized}. In particular, quantum circuit models can exploit interference and entanglement to represent high-order correlated probability distributions that are challenging for classical models to capture, motivating their use as learnable generative priors~\cite{liu2018differentiable,benedetti2019generative}.

However, access to an exponentially large Hilbert space alone does not guarantee a practical advantage over classical models; such advantages depend on the learning task, data encoding, and available computational resources~\cite{cerezo2022challenges,huang2021power,schuld2021encoding}. In the current Noisy Intermediate-Scale Quantum (NISQ) era~\cite{preskill2018quantum}, applying QML to large-scale, real-world data remains challenging. Many hybrid quantum-classical architectures optimize classical and quantum parameters jointly, with classical updates depending on derivatives of quantum expectation values. Parameter-shift methods provide analytic gradient formulas, but their evaluation on hardware requires repeated circuit executions and measurements, creating substantial overhead as the number of trainable parameters and required sampling precision increase~\cite{schuld2019evaluating,sweke2020stochastic}. Independently of this gradient coupling, trainability can be limited by ``barren plateaus'': gradient variances may decrease exponentially with system size for certain circuit ensembles and cost functions~\cite{mcclean2018barren,cerezo2021cost}, and hardware noise can induce further exponential gradient suppression under appropriate depth-scaling conditions~\cite{wang2021noise}. Encoding high-dimensional classical inputs into quantum states also incurs state-preparation costs that can offset potential computational gains~\cite{biamonte2017quantum,cerezo2022challenges,schuld2021encoding}. Finally, models that retain quantum circuits in their prediction pipeline require quantum execution and repeated measurements at deployment~\cite{benedetti2019parameterized}, making hardware access and execution latency important constraints for real-time applications.

To circumvent these QML challenges in the NISQ era, recent literature has developed a range of mitigation strategies targeting optimization landscapes, gradient estimation costs, and data encoding. For barren plateaus, proposed mitigations include adopting local cost functions~\cite{cerezo2021cost}, layer-wise parameter updates~\cite{skolik2021layerwise}, and identity-block initialization~\cite{grant2019initialization}---all aiming to maintain non-zero gradient variances during shallow or layered training. To reduce the prohibitive cost of gradient estimation, stochastic gradient descent with drastically reduced measurement shots~\cite{sweke2020stochastic} and gradient-free Simultaneous Perturbation Stochastic Approximation (SPSA)~\cite{spall1992multivariate} are advocated, effectively bypassing the parameter-shift rule's heavy circuit evaluations. For the data-loading bottleneck, transfer learning~\cite{mari2020transfer} leverages classical pre-trained networks to compress high-dimensional inputs into low-dimensional features before quantum encoding, while data re-uploading~\cite{perez2020data} processes input features sequentially within a fixed small-qubit circuit, avoiding massive parallel state preparation. However, these strategies are mostly validated under small-scale idealized simulations, often assume instantaneous classical--quantum interaction, and overlook communication overhead. Moreover, the encoding schemes either rely on pre-trained classical models that may dominate outcomes or suffer from saturation in expressive power with growing data dimensionality~\cite{schuld2021encoding,sim2019expressibility}.

To mitigate the real-time quantum execution latency incurred when a QML model must access quantum hardware at inference time, existing studies have explored several routes for moving quantum computation offline or replacing online quantum execution with classical prediction. One line of work focuses on quantum-to-classical model extraction, where a trained quantum model is converted into a classical surrogate for deployment~\cite{schreiber2023surrogates}. For example, classical-shadow techniques can be used to construct a ``shadow model'' after the quantum training phase so that subsequent predictions can be evaluated classically~\cite{huang2024shadows}. Another line of work can be viewed as quantum-assisted classical learning, where quantum computation is used during training to generate auxiliary learning signals, such as quantum kernels, relational structures, or training updates, while the deployed predictor remains classical. Representative examples include quantum-enhanced feature-space methods that use quantum kernels with classical kernel learners~\cite{havlicek2019supervised}, quantum relational knowledge distillation that injects quantum relational information into the training of a classical student model~\cite{liu2025quantum}, and quantum-assisted training schemes for classical neural networks~\cite{du2024training}. However, surrogate-based extraction is typically a post-hoc step restricted to model families with tractable classical descriptions, whereas quantum-assisted approaches still require quantum processing of the training data itself, through kernel estimation, relational knowledge extraction, or quantum-computed training signals. As a result, these approaches shift the quantum computational burden from online inference to offline stages rather than eliminating it, and the associated measurement and circuit-execution costs continue to scale with the dataset size and input dimensionality.

A recent work by Chen~\cite{chen2026hybrid} addresses these bottlenecks by combining an LSTM~\cite{hochreiter1997lstm} with a Quantum Circuit Born Machine (QCBM)~\cite{liu2018differentiable,benedetti2019generative} as a learnable, unconditional generative prior. A stochastic ``Drop-Prior'' mechanism randomly disables the prior during training, encouraging implicit distillation of its structural information into the LSTM without a separate surrogate-model extraction stage. This approach is conceptually aligned with implicit knowledge distillation through indirect learning signals~\cite{hinton2015distilling,gou2021distillation,wang2023explicit,xue2024icd}. The unconditional QCBM generates prior samples independently of the input, avoiding direct encoding of high-dimensional classical data into quantum states, while alternating optimization removes end-to-end quantum--classical gradient coupling and the associated parameter-shift evaluations. The trained LSTM can then perform fully classical inference with the QCBM disabled, eliminating online quantum execution, although quantum sampling and optimization remain necessary during training. However, this work has only been applied to time-series forecasting and evaluated with a single classical backbone, LSTM, leaving its applicability to other tasks and backbone architectures unexplored.

Building on the QCBM-based approach of Chen~\cite{chen2026hybrid}, we develop \textbf{QPAT}, a unified framework that extends quantum-assisted training beyond LSTM-based financial volatility forecasting to diverse prediction tasks and classical architectures. The framework retains the original alternating, fitness-guided prior learning and stochastic ``Drop-Prior'' mechanism, while formalizing a common interface for integrating quantum priors with task-specific classical representations. Our contribution centers on this unified formulation, its adaptation to regression and classification, and a systematic investigation of whether the benefits of prior-assisted training and fully classical deployment extend across modalities and backbone architectures. We complement these evaluations with classical-prior ablations, qualitative case studies, and a mechanism-oriented analysis connecting Drop-Prior to implicit distillation.

The main contributions of this work are summarized as follows:
\begin{itemize}
    \item \textbf{Unified formulation and cross-domain evaluation.} We generalize the LSTM-based implementation into a modular architecture comprising classical representation extraction, prior projection and fusion, and a task-specific prediction head. This formulation supports both regression and classification and provides a common interface for adapting the QCBM prior to different backbone architectures. On this basis, we extend the empirical investigation beyond financial volatility forecasting to additional time series, image classification, and text classification tasks, using LSTM, ResNet-18~\cite{he2016resnet}, and DistilBERT~\cite{sanh2019distilbert} as representative recurrent, convolutional, and Transformer backbones, and examine whether the benefits of the original approach extend beyond its initial task and architecture.
    \item \textbf{Performance gains and ablation evidence.} We show that, across all six benchmarks, models trained with the QCBM prior consistently outperform their purely classical counterparts and exhibit lower performance variance across random seeds. Grad-CAM~\cite{selvaraju2017gradcam} and Integrated Gradients~\cite{sundararajan2017axiomatic} case studies provide complementary qualitative evidence about model predictions and attribution patterns. We further compare QCBM priors with uniform and RBM priors to assess whether these gains can be reproduced by unstructured randomness or the tested classical generative alternative, and find that neither classical alternative matches the QCBM prior.
    \item \textbf{Cross-task validation of classical deployment.} We compare quantum--classical and classical inference for the same trained hybrid models across tasks, assessing whether performance gains persist after the quantum prior is removed. By verifying that accuracy is preserved without quantum execution at deployment, this evaluation establishes fully classical inference as a general property of prior-assisted hybrid training across modalities.
    \item \textbf{Mechanism-oriented analysis.} We examine the expected Drop-Prior training objective and its local regularization effects, and develop conceptual connections to implicit distillation. This analysis offers an interpretive account of how a removable prior reshapes the local optimization landscape of the classical predictor, complementing the empirical evidence across tasks.
\end{itemize}

\section{Methodology}

\subsection{QCBM-based Quantum Generative Prior}
The QCBM serves as the foundational core of our quantum prior module. Unlike classical generative models such as Variational Autoencoders (VAEs)~\cite{kingma2014vae} or Generative Adversarial Networks (GANs)~\cite{goodfellow2014gan}, a QCBM represents a probability distribution directly through the wave function of a parameterized quantum circuit (PQC)~\cite{benedetti2019parameterized,sim2019expressibility}. Given a circuit with parameters $\boldsymbol{\theta}$, the probability of measuring a specific bitstring $e \in \{0, 1\}^n$ follows the Born rule:
\begin{equation}
    P_{\boldsymbol{\theta}}(e) = |\langle e | \psi(\boldsymbol{\theta}) \rangle|^2.
\end{equation}
Operating in a Hilbert space whose dimension grows exponentially with the qubit count, the QCBM can compactly represent distributions with high-order correlations over bitstrings, which motivates its use as a structured generative prior in place of classical alternatives of comparable parameter count. Crucially, the QCBM acts as an independent, unconditional distribution generator. Its operational space is defined by the number of qubits and the circuit depth, completely independent of the input classical data dimensions. This decoupling allows our framework to entirely avoid the quantum data-encoding bottleneck that affects traditional QML models.

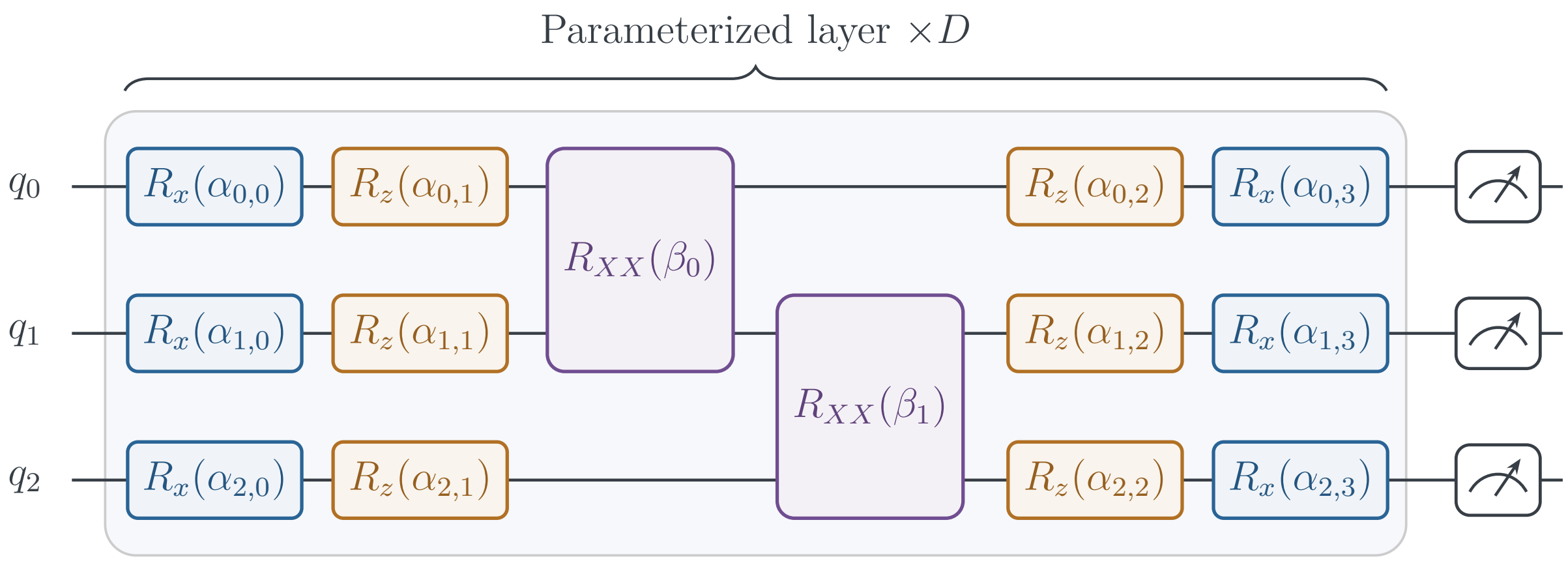
\begin{figure}[htbp]
    \centering
    \resizebox{\linewidth}{!}{%
        \begin{tikzpicture}[
            x=1cm,
            y=1cm,
            font=\large,
            line width=0.85pt,
            wire/.style={draw=wirecolor},
            rotation/.style={
                rectangle,
                rounded corners=3pt,
                minimum width=1.78cm,
                minimum height=0.78cm,
                inner sep=3pt,
                line width=1pt
            },
            rx/.style={
                rotation,
                draw=rxcolor,
                fill=rxcolor!7,
                text=rxcolor!85!black
            },
            rz/.style={
                rotation,
                draw=rzcolor,
                fill=rzcolor!8,
                text=rzcolor!85!black
            },
            entangler/.style={
                rectangle,
                rounded corners=5pt,
                draw=xxcolor,
                fill=xxcolor!8,
                text=xxcolor!85!black,
                minimum width=1.90cm,
                minimum height=2.28cm,
                inner sep=3pt,
                line width=1pt
            }
        ]
        \begin{scope}[on background layer]
            \filldraw[
                fill=panelcolor,
                draw=black!20,
                rounded corners=9pt,
                line width=0.7pt
            ] (-0.12,0.77) rectangle (13.18,-3.77);
        \end{scope}

        \draw[
            decorate,
            decoration={brace,amplitude=7pt},
            draw=wirecolor,
            line width=0.8pt
        ] (0.08,0.98) -- (12.98,0.98);
        \node[font=\large,text=wirecolor,anchor=south]
            at (6.53,1.24) {Parameterized layer \(\times D\)};

        \foreach \j/\yy in {0/0,1/-1.5,2/-3.0}{
            \node[anchor=east,text=wirecolor]
                at (-0.62,\yy) {\(q_{\j}\)};
            \draw[wire] (-0.46,\yy) -- (14.78,\yy);
        }

        \foreach \j/\yy in {0/0,1/-1.5,2/-3.0}{
            \node[rx] at (1.00,\yy) {\(R_x(\alpha_{\j,0})\)};
            \node[rz] at (3.10,\yy) {\(R_z(\alpha_{\j,1})\)};
        }

        \node[entangler] at (5.35,-0.75) {\(R_{XX}(\beta_0)\)};
        \node[entangler] at (7.70,-2.25) {\(R_{XX}(\beta_1)\)};

        \foreach \j/\yy in {0/0,1/-1.5,2/-3.0}{
            \node[rz] at (10.00,\yy) {\(R_z(\alpha_{\j,2})\)};
            \node[rx] at (12.10,\yy) {\(R_x(\alpha_{\j,3})\)};
        }

        \foreach \yy in {0,-1.5,-3.0}{
            \begin{scope}[shift={(14.12,\yy)}]
                \draw[
                    draw=wirecolor,
                    fill=white,
                    rounded corners=4pt,
                    line width=0.9pt
                ] (-0.43,-0.36) rectangle (0.43,0.36);
                \draw[draw=wirecolor]
                    (-0.29,-0.12)
                    .. controls (-0.16,0.12) and (0.16,0.12) ..
                    (0.29,-0.12);
                \draw[
                    draw=wirecolor,
                    -{Stealth[length=4.5pt,width=3.5pt]}
                ] (-0.03,-0.16) -- (0.23,0.22);
            \end{scope}
        }
        \end{tikzpicture}%
    }
    \caption{A three-qubit example of the QCBM architecture.
    The shaded block represents a parameterized layer repeated
    \(D\) times, with measurements performed after the final
    layer. The same construction extends to \(n\) qubits by
    applying local rotations to each qubit and sequential
    \(R_{XX}\) gates to nearest-neighbor pairs.}
    \label{fig:qcbm}
\end{figure}

In this work, we employ a QCBM comprising \(n\) qubits and \(D\) parameterized layers. Figure~\ref{fig:qcbm} presents a three-qubit example of this architecture. Each layer begins with local rotations: an \(R_x(\alpha_{j,0})\) gate followed by an \(R_z(\alpha_{j,1})\) gate is applied to every qubit \(q_j\), where \(j=0,\ldots,n-1\). Next, nearest-neighbor correlations are introduced through \(R_{XX}(\beta_j)\) gates acting on the pairs \((q_j,q_{j+1})\). These entangling gates are applied sequentially in increasing order of \(j\), from \(j=0\) to \(n-2\). Each layer ends with an \(R_z(\alpha_{j,2})\) rotation followed by an \(R_x(\alpha_{j,3})\) rotation on every qubit. The layer structure is repeated \(D\) times, after which all \(n\) qubits are measured in the computational basis to obtain an \(n\)-bit classical sample. Both the single-qubit rotation parameters \(\alpha\) and the two-qubit interaction parameters \(\beta\) are trainable. By optimizing these parameters, the QCBM learns to approximate a target prior distribution. Samples from the learned distribution are generated by repeatedly executing and measuring the circuit.

\subsection{Generalized Architecture}
\begin{figure}[htbp!]
    \centering
    \includegraphics[width=\linewidth]{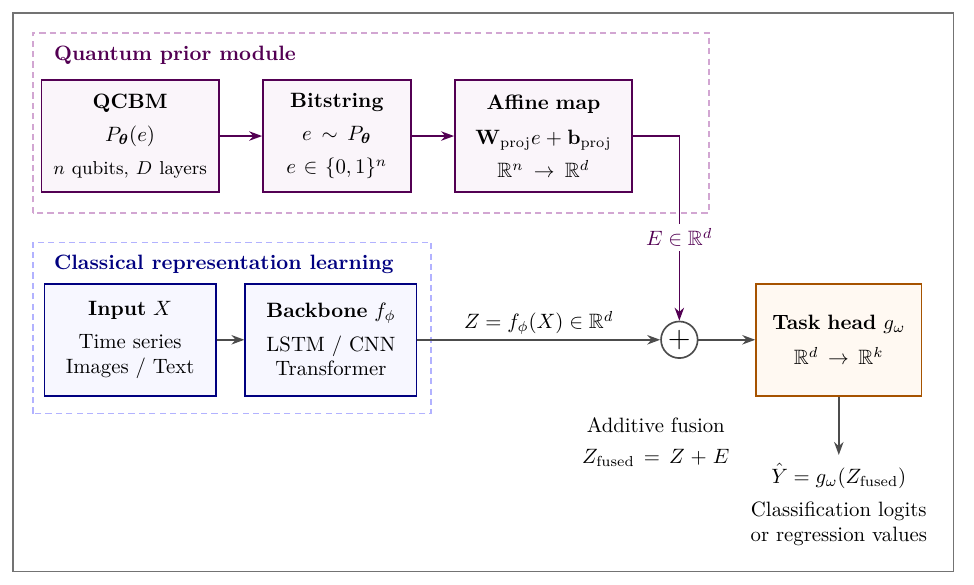}
    \caption{Generalized architecture of the QPAT framework.
    A modality-specific classical backbone maps the input \(X\) to
    a feature vector \(Z\in\mathbb R^d\). Independently of the input,
    the QCBM generates a bitstring \(e\), which is mapped to a prior
    embedding \(E\in\mathbb R^d\) by a trainable affine projection.
    Element-wise addition produces the fused representation \(Z+E\),
    and the task head \(g_\omega\) produces the prediction
    \(\hat{Y}=g_\omega(Z+E)\), representing classification logits or regression values.}
    \label{fig:qcbm-prior-architecture}
\end{figure}

The QPAT framework is designed to be highly modular, decoupling task-specific representation learning from the generalized structured inductive bias provided by the quantum prior. It consists of three interconnected components: a classical feature extractor, a quantum prior embedding module, and a modality-agnostic fusion mechanism, as illustrated in Figure~\ref{fig:qcbm-prior-architecture}.

\textbf{(i) Classical Backbone ($f_\phi$):} Let $\mathcal{X}$ denote the input space, which varies strictly by modality (e.g., $\mathcal{X} = \mathbb{R}^{H \times W \times C}$ for images with height $H$, width $W$, and $C$ channels, $\mathcal{X} = \mathbb{R}^{T \times F}$ for multivariate time series with $T$ time steps and $F$ variates, or discrete token sequences for text). The classical backbone acts as a universal, modality-specific feature extractor:
\begin{equation}
    Z = f_\phi(X), \quad \text{where} \quad f_\phi: \mathcal{X} \rightarrow \mathbb{R}^d.
\end{equation}
Rather than being constrained to a particular neural architecture, $f_\phi$ serves as an interchangeable plug-in module designed to capture domain-specific inductive biases---such as spatial invariance~\cite{he2016resnet}, semantic self-attention~\cite{vaswani2017attention}, or temporal dependencies. It projects the high-dimensional input $X \in \mathcal{X}$ into a dense, abstract hidden representation $Z$ embedded in a $d$-dimensional latent space.

\textbf{(ii) Quantum Prior Module:} On the quantum side, the QCBM generates a discrete prior. Let $e \in \{0, 1\}^n$ be a bitstring sampled from the trained quantum distribution $P_{\boldsymbol{\theta}}(e)$. Because $e$ exists in an $n$-dimensional Boolean hypercube, it must be mapped to the continuous $d$-dimensional latent space of the classical backbone. This is achieved via a trainable affine projection layer:
\begin{equation}
    E= \mathbf{W}_{proj} \cdot e + \mathbf{b}_{proj}
\end{equation}
where $\mathbf{W}_{proj} \in \mathbb{R}^{d \times n}$ is the projection weight matrix, $\mathbf{b}_{proj} \in \mathbb{R}^d$ is the bias vector, and $E\in \mathbb{R}^d$ represents the dense quantum prior embedding.

\textbf{(iii) Modality-Agnostic Fusion and Task Head:} To integrate the classical inductive bias with the quantum structural prior, the two representations are combined using a modality-agnostic fusion operation. In its simplest form, this is implemented as an element-wise addition:
\begin{equation}
    Z_{fused} = Z + E.
\end{equation}
Finally, the fused representation is fed into a task-specific head $g_\omega$, which maps the $d$-dimensional latent vector to the output space $\mathbb{R}^k$ (e.g., pre-softmax logits for classification or continuous values for regression), with any output nonlinearity such as the softmax absorbed into the task loss $\mathcal{L}_{task}$:
\begin{equation}
    \hat{Y} = g_\omega(Z_{fused}), \quad \text{where} \quad g_\omega: \mathbb{R}^d \rightarrow \mathbb{R}^k.
\end{equation}
This architectural formulation ensures that the quantum module remains completely agnostic to the input data dimensionality, thereby circumventing the data encoding bottleneck entirely.

\subsection{Universal Alternating Training}
To eliminate the severe gradient coupling between the quantum and classical modules, we employ a decoupled, alternating training strategy.

\textbf{Phase 1 (Classical Training):} The parameters $\boldsymbol{\theta}$ of the QCBM are frozen. We first sample a set of prior bitstrings from the QCBM, equal in size to the training set, so that each training sample is assigned one bitstring $e$, and compute the corresponding embedding $E$. During this phase, the prior is stochastically activated via the Drop-Prior mechanism detailed in Section~\ref{sec:Drop-Prior Mechanism}. The classical backbone $f_\phi$, the projection layer, and the task head $g_\omega$ are then updated via standard backpropagation to minimize the expected task loss over both the sampled priors and the stochastic mask:
\begin{equation}
\min_{\phi, \mathbf{W}_{proj}, \mathbf{b}_{proj}, \omega} \; \mathbb{E}_{(X,Y)\sim\mathcal{D}} \, \mathbb{E}_{e \sim P_{\boldsymbol{\theta}}} \, \mathbb{E}_{\delta \sim \mathrm{Bernoulli}(p)} \left[ \mathcal{L}_{task}\!\left( g_{\omega}\!\left( f_{\phi}(X) + \delta\,E \right), Y \right) \right],
\end{equation}
where $E = \mathbf{W}_{proj}\, e + \mathbf{b}_{proj}$ is the projected embedding of the sampled bitstring, $\delta$ is the Drop-Prior mask, and $\mathcal{L}_{task}$ denotes a universal task loss (e.g., Cross-Entropy for classification tasks, Mean Squared Error for regression tasks).

\textbf{Phase 2 (QCBM Optimization):} The classical parameters $\phi$, $\mathbf{W}_{proj}$, $\mathbf{b}_{proj}$, and $\omega$ are frozen. To guide the QCBM towards generating useful priors, we evaluate a generalized fitness function for each uniquely sampled bitstring $e_j$:
\begin{equation}
    S(e_j) = -\mathcal{L}_{task}(\hat{Y}, Y \mid e_j),
\end{equation}
where $\hat{Y}$ is the prediction conditioned on the prior $e_j$. These scores are transformed into a target probability distribution $P_{target}$ via a softmax function. The QCBM parameters $\boldsymbol{\theta}$ are then updated to minimize the Kullback-Leibler (KL) divergence between its generated Born distribution $P_{\boldsymbol{\theta}}$ and the target distribution $P_{target}$.

This decoupled training mechanism offers three practical advantages over conventional hybrid architectures. First, by eliminating gradient backpropagation through the quantum module, it minimizes the interaction between the quantum and classical modules during optimization. In contrast to tightly coupled hybrid models, where chained gradient computations require frequent quantum-classical communication, our approach substantially reduces the training overhead and improves efficiency, a benefit that is particularly pronounced on real quantum hardware. Second, since the optimization of the QCBM does not rely on gradient information from the classical module, it can be carried out with gradient-free algorithms such as COBYLA, which not only accelerate the convergence of the quantum parameters but also exhibit stronger noise resilience than gradient-based optimizers on current NISQ devices. Third, the framework eliminates the data-loading bottleneck. Instead of encoding large-scale classical data into quantum states, a process whose circuit complexity can grow rapidly with the data dimension and offset the potential advantages of quantum computing, the QCBM acts purely as a generator that learns a prior distribution, so classical data never needs to be loaded onto the quantum device.

\subsection{Drop-Prior Mechanism and Implicit Distillation}
\label{sec:Drop-Prior Mechanism}
While the alternating optimization scheme described above reduces the \emph{training} cost of the hybrid framework, the \emph{inference} cost would remain substantial whenever the quantum module must be queried at deployment. To enable low-cost, fully classical inference, we adopt the stochastic \textit{Drop-Prior} mechanism first proposed in~\cite{chen2026hybrid}, which facilitates implicit knowledge distillation~\cite{hinton2015distilling,gou2021distillation} and is conceptually related to stochastic regularization techniques such as dropout~\cite{srivastava2014dropout} and stochastic depth~\cite{huang2016stochastic}. Whereas the original work introduced this strategy purely empirically, we formalize its effect on the classical model through the following analysis. Consider a single sample $(X,Y)$. The classical backbone $f_{\phi}$ maps the input $X$ to a deterministic feature vector $Z = f_{\phi}(X) \in \mathbb{R}^d$. A quantum prior embedding $E$ is sampled from the trained QCBM, and a Bernoulli random variable $\delta \sim \mathrm{Bernoulli}(p)$ controls whether the prior branch is activated. Let the student prediction (prior disabled) be $\hat{Y}_{\text{without}} = g_{\omega}(Z)$, and the teacher prediction (prior enabled) be $\hat{Y}_{\text{with}} = g_{\omega}(Z+E)$, where $g_{\omega}: \mathbb{R}^d \to \mathbb{R}^k$ is the prediction head and $k$ denotes the output dimension (e.g., the number of logits for classification tasks).

To quantify how the teacher path influences the student path, we denote the teacher loss by $\ell_t(E) := \ell(\hat{Y}_{\text{with}}, Y) = \ell(g_{\omega}(Z+E), Y)$, which is a composite function of the prior $E$, and the student loss by $\ell_s := \ell(\hat{Y}_{\text{without}}, Y) = \ell_t(0)$. Here the loss $\ell \equiv \mathcal{L}_{task}$ denotes the universal task loss (e.g., cross-entropy for classification or mean squared error for regression) regarded as a function of the prediction $\hat{Y}$. We directly expand $\ell_t(E)$ around $E = 0$. Assuming $\ell$ is smooth and at least twice differentiable with respect to its prediction, a second-order Taylor series gives
\begin{equation}
\ell_t(E) = \ell_s + \nabla_E \ell_s^\top E + \frac{1}{2} E^\top \nabla_E^2 \ell_s\, E + \mathcal{O}(\|E\|^3).
\end{equation}
By the chain rule, the derivatives of the composite function $\ell_t$ at $E = 0$ are
\begin{equation}
\nabla_E \ell_s = J_g^\top\, \nabla_{\hat{Y}}\ell_s, \qquad
\nabla_E^2 \ell_s = J_g^\top H J_g + \sum_{i=1}^{k} (\nabla_{\hat{Y}}\ell_s)_i\, \nabla_Z^2 g_{\omega,i},
\end{equation}
where $J_g = \partial g_{\omega}/\partial Z$ is the (constant) $k \times d$ Jacobian matrix of the prediction head with respect to the feature $Z$, and $H = \nabla_{\hat{Y}}^2\ell_s = \nabla_{\hat{Y}}^2\ell(\hat{Y}_{\text{without}}, Y)$ is the Hessian of the student loss with respect to the prediction. Since $g_{\omega}$ is affine in the feature $Z$, the head curvature term vanishes identically, $\nabla_Z^2 g_{\omega,i} = 0$ for $i = 1, \dots, k$, and the composite Hessian reduces exactly to the Gauss--Newton term $J_g^\top H J_g$. Substituting these derivatives into the expansion and applying the trace identity $E^\top J_g^\top H J_g E = \operatorname{Tr}\!\left(J_g^\top H J_g\, E E^\top\right)$ yields the second-order approximation of the teacher loss:
\begin{equation} \label{SOA}
\ell_t(E) = \ell_s + \tilde{\mathcal{T}}_1 + \frac{1}{2}\tilde{\mathcal{T}}_2 + \mathcal{O}(\|E\|^3),
\end{equation}
with the per-sample auxiliary quantities defined as
\begin{equation}
\tilde{\mathcal{T}}_1 = \nabla_{\hat{Y}}\ell_s^\top J_g E, \qquad
\tilde{\mathcal{T}}_2 = \operatorname{Tr}\left(J_g^\top H J_g \cdot E E^\top\right).
\end{equation}

The instantaneous loss is written explicitly according to the mask $\delta$:
\begin{equation}
\mathcal{L}_{\text{inst}} = \delta \cdot \ell_t + (1 - \delta) \cdot \ell_s.
\end{equation}
Substituting the teacher approximation in \eqref{SOA} and rearranging terms gives a concise linear expression for the instantaneous loss in terms of the random mask:
\begin{equation}
\mathcal{L}_{\text{inst}} = \ell_s + \delta \cdot \left( \tilde{\mathcal{T}}_1 + \frac{1}{2}\tilde{\mathcal{T}}_2 \right) + \mathcal{O}(\|E\|^3).
\end{equation}
Taking the expectation over $\delta$ and using $\mathbb{E}[\delta] = p$, we obtain the conditional expectation
\begin{equation}
\mathbb{E}_{\delta}[\mathcal{L}_{\text{inst}}] = \ell_s + p\tilde{\mathcal{T}}_1 + \frac{p}{2}\tilde{\mathcal{T}}_2 + \mathcal{O}(\|E\|^3).
\end{equation}
Finally, taking the global expectation over all random variables $(X,Y,E)$ and defining the population quantities $\mathcal{T}_1 = \mathbb{E}[\tilde{\mathcal{T}}_1]$ and $\mathcal{T}_2 = \mathbb{E}[\tilde{\mathcal{T}}_2]$, we arrive at the second-order expansion of the expected loss under the Drop-Prior mechanism:
\begin{equation} \label{SOE}
\mathcal{L}(p) = \mathbb{E}[\ell_s] + p \cdot \mathcal{T}_1 + \frac{p}{2} \cdot \mathcal{T}_2 + \mathcal{O}(\|E\|^3).
\end{equation}

Based on \eqref{SOE}, we now examine the physical roles of $\mathcal{T}_1$ and $\mathcal{T}_2$, which play distinct but coupled roles. For the first-order term, the Jacobian $J_g$ is constant for the default linear prediction head, and the QCBM sample $E$ is independent of the data pair $(X,Y)$, so the joint expectation factorizes exactly as $\mathcal{T}_1 = \bar{g}^\top J_g \mu_E$, where $\bar{g} := \mathbb{E}_{X,Y}[\nabla_{\hat{Y}}\ell_s]$ denotes the mean loss gradient and $\mu_E = \mathbb{E}_E[E]$. Since $\mathcal{T}_1$ is a signed scalar, the optimizer may drive it negative, thereby lowering the training loss through the prior pathway. This behavior can be viewed as a form of shortcut learning~\cite{geirhos2020shortcut}: the backbone $f_{\phi}$ is not forced to extract meaningful features from the input $X$; instead, it can degenerate into an accessory to the quantum prior, while $\mathbb{E}[\ell_s]$ itself remains poorly optimized.

The second-order term $\mathcal{T}_2 = \operatorname{Tr}\left(\mathbb{E}_{X,Y}\left[J_g^\top H J_g\right] M_E\right)$ plays the complementary role, where $M_E := \mathbb{E}_E[E E^\top] = \Sigma_E + \mu_E\mu_E^\top$ denotes the second moment of the prior embedding, with mean $\mu_E = \mathbb{E}_E[E]$ and covariance $\Sigma_E$; here we have again used the independence between the QCBM sample $E$ and the data pair $(X,Y)$ to factorize the joint expectation. For convex losses we have $H \succeq 0$; in particular, for the cross-entropy loss with softmax, $H = \operatorname{diag}(\hat{\pi}) - \hat{\pi}\hat{\pi}^\top$ is the covariance matrix of the predicted categorical distribution $\hat{\pi} := \mathrm{softmax}(\hat{Y})$ and hence positive semi-definite. Since $M_E \succeq 0$ holds unconditionally (as the sum of the covariance and the rank-one mean term), it follows that $\mathcal{T}_2 \ge 0$. This term measures the projection of the expected loss Hessian onto the second-moment directions of the quantum prior, which comprise both the fluctuation directions encoded in $\Sigma_E$ and the systematic bias direction $\mu_E$ learned by the QCBM; it penalizes the sensitivity of the prediction head $g_{\omega}$ to feature perturbations $E$. A smaller value of $\mathcal{T}_2$ indicates a flatter loss landscape and thus greater robustness to removing the prior at test time. Hence $\mathcal{T}_2$ acts as a benign regularizer that promotes generalization.

When $0 < p < 1$, the random dropping strategy exerts two synergistic effects. First, it directly weakens the strength of shortcut learning: because the coefficient of $\mathcal{T}_1$ in \eqref{SOE} is reduced from $1$ (at $p=1$) to $p$, the contribution of negative co-adaptation to the total loss is suppressed, forcing the backbone to rely, to some extent, on its own capacity to reduce $\mathbb{E}[\ell_s]$. Second, and more importantly, the random mask introduces a strong stochastic regularization effect arising from the first-order fluctuations of the instantaneous loss~\cite{bishop1995noise,wager2013dropout}. To isolate the noise injected by the dropping mechanism itself from the inherent sample-to-sample fluctuations of the data, we condition on the sample $(X, Y, E)$ and consider only the fluctuation of the loss induced by the random mask. Let the total contribution of the teacher path in a single forward pass be
\begin{equation}
A = \tilde{\mathcal{T}}_1 + \frac{1}{2}\tilde{\mathcal{T}}_2.
\end{equation}
With $A$ defined above, the instantaneous loss at second order is $\mathcal{L}_{\text{inst}} = \ell_s + \delta A$, where both $\ell_s$ and $A$ are constant with respect to the mask once the sample $(X, Y, E)$ is given. Its conditional mean is $\mathbb{E}_{\delta}[\mathcal{L}_{\text{inst}}] = \ell_s + pA$, so the fluctuation around this mean is $\mathcal{L}_{\text{inst}} - \mathbb{E}_{\delta}[\mathcal{L}_{\text{inst}}] = (\delta - p)\,A$. Since $\delta \in \{0, 1\}$ is Bernoulli with $\mathbb{E}[\delta^2] = \mathbb{E}[\delta] = p$, its variance is $\operatorname{Var}(\delta) = p - p^2 = p(1-p)$, and the conditional variance (second central moment) of the instantaneous loss induced by the random mask is
\begin{equation}
\operatorname{Var}_{\delta}\big(\mathcal{L}_{\text{inst}} \,\big|\, X, Y, E\big) = \mathbb{E}_{\delta}\big[\big((\delta - p)\,A\big)^2\big] = A^2\,\mathbb{E}_{\delta}\big[(\delta - p)^2\big] = A^2\,\operatorname{Var}(\delta) = p(1-p)\,A^2.
\end{equation}
The conditional variance $p(1-p)\,A^2$ reaches its maximum $\frac{1}{4}\,A^2$ at $p = 0.5$, meaning that the fluctuations injected by the dropping mechanism are most severe at this point. In stochastic gradient-based optimization, such fluctuations translate into strong implicit gradient noise. According to stochastic differential equation theory, higher noise intensity drives the model parameters away from sharp minima and forces convergence to flat minima~\cite{keskar2017sharp,chaudhari2017entropy}. Flat minima are inherently robust to small perturbations in the feature space; consequently, even if the quantum prior $E$ is completely removed during inference, the model output does not deviate significantly. This is the mathematical origin of the implicit knowledge distillation effect.

The extreme case $p=1$ offers a clear contrast. When the quantum prior branch is never dropped, the expected loss in \eqref{SOE} reduces to
\begin{equation}
\mathcal{L}(1) = \mathbb{E}[\ell_s] + \mathcal{T}_1 + \frac{1}{2}\mathcal{T}_2.
\end{equation}
Here, the coefficient of $\mathcal{T}_1$ is restored to unity, so the optimizer can fully exploit negative co-adaptation and tends to rely heavily on the prior $E$. The backbone's incentive to reduce $\mathbb{E}[\ell_s]$ through its own features is thereby weakened, and a large value of $\mathbb{E}[\ell_s]$ can remain concealed during training by the contribution of $E$. Once $E$ is removed at inference, this concealed error is exposed. Moreover, when $p=1$, the mask-induced variance vanishes:
\begin{equation}
\operatorname{Var}_{\delta}\big(\mathcal{L}_{\text{inst}} \,\big|\, X, Y, E\big) = 1 \cdot 0 \cdot A^2 = 0.
\end{equation}
Thus, although the ordinary mini-batch sampling noise is still present, the additional gradient noise injected by the dropping mechanism vanishes, and the model no longer benefits from its regularizing effect. Without this noise-driven bias toward flat regions, the optimizer is more inclined to settle into a sharper minimum, where the feature-space curvature $J_g^\top H J_g$ exhibits larger eigenvalues. At such a minimum, the model becomes highly sensitive to perturbations of $E$: removing the quantum prior at test time is equivalent to applying a deterministic perturbation $-E$ to the feature $Z$, and according to the residual second-order term $\frac{1}{2}E^\top J_g^\top H J_g\, E$ in the Taylor expansion, this perturbation can induce a substantial quadratic increase in the loss. The concealed cost of $\mathbb{E}[\ell_s]$ (previously masked by $\mathcal{T}_1$) and this heightened sensitivity to the removal of $E$ jointly account for the pronounced performance drop.

\section{Experiments and Results}
To evaluate the generality and robustness of the QPAT framework, we conducted experiments across three machine learning domains: time-series forecasting, computer vision, and natural language processing. All experiments were repeated across three independent random seeds, with results reported as mean $\pm$ standard deviation.

\subsection{Setup}

\subsubsection{Datasets}
\label{sec:datasets}

We evaluate our method on six datasets spanning three tasks:
time-series forecasting, image classification, and text classification.
Table~\ref{tab:dataset_splits} summarizes the sizes of the training,
validation, and test partitions.

\begin{itemize}
    \item \textbf{Time-series forecasting.}
    We use the Exchange~\cite{lai2018lstnet} and influenza-like illness (ILI)~\cite{wang2026deep} datasets.
    Exchange contains daily exchange-rate series for eight currencies
    from 1990 to 2016.
    ILI contains weekly influenza-like illness observations
    reported by the US Centers for Disease Control and Prevention (CDC)
    from 2002 to 2021.
    For these two datasets, the partition sizes in
    Table~\ref{tab:dataset_splits} refer to the numbers of time steps
    before sliding-window construction.

    \item \textbf{Image classification.}
    We use a binary classification subset of the Breast Ultrasound Images (BUSI) dataset~\cite{aldhabyani2020busi}
    and a subset of the Aircraft dataset~\cite{maji2013aircraft}.
    The BUSI subset contains breast ultrasound images and is used for the diagnostic classification of tumors as either benign or malignant.
    The Aircraft subset comprises eight Boeing 737 variants,
    ranging from the 737-200 to the 737-900, and is used for
    fine-grained aircraft classification.

    \item \textbf{Text classification.}
    We use the Recognizing Textual Entailment (RTE)~\cite{wang2018glue} and
    Text REtrieval Conference (TREC)~\cite{li2002trec} datasets.
    RTE is a sentence-pair classification task that determines
    whether a premise entails a hypothesis.
    We use the officially provided labeled validation set as our
    test set and split the official training set into training
    and validation partitions at an 80:20 ratio.
    The resulting training, validation, and test sets contain
    1,992, 498, and 277 sentence pairs, respectively.
    TREC is a fine-grained question classification benchmark; in our experiments, we adopt a 47-category configuration. This setting is considerably more challenging than coarse-grained classification due to the scarcity of training examples per category, severe class imbalance, and the inherently ambiguous boundaries between fine-grained categories.
\end{itemize}

\begin{table}[t]
    \centering
    \caption{Dataset partition sizes.
    For time-series forecasting, counts indicate time steps
    before sliding-window construction; for image and text
    classification, counts indicate images and text examples,
    respectively.}
    \label{tab:dataset_splits}
    \small
    \setlength{\tabcolsep}{4pt}
    \begin{tabular}{llrrr}
        \toprule
        Task & Dataset & Training & Validation & Test \\
        \midrule
        Time-series forecasting & Exchange & 5,311 & 759   & 1,518 \\
                                & ILI  & 676   & 96    & 194   \\
        \midrule
        Image classification    & BUSI     & 440   & 94    & 95    \\
                                & Aircraft & 267   & 266   & 267   \\
        \midrule
        Text classification     & RTE      & 1,992 & 498   & 277   \\
                                & TREC     & 4,361 & 1,091 & 500   \\
        \bottomrule
    \end{tabular}
\end{table}
\subsubsection{Models}
For each task, we compare a pure classical backbone against our Drop-Prior Hybrid model, which augments the classical predictor with an additional QCBM module. During training, the Drop-Prior mechanism randomly disables the quantum prior module with probability $1-p$, encouraging the classical network to absorb the prior-induced structural signal through implicit distillation. At evaluation time, the same trained Drop-Prior Hybrid model is assessed under \textit{Quantum--Classical inference}, with the quantum prior module enabled, and \textit{Classical inference}, with the quantum prior completely removed and predictions based solely on the classical parameters learned during hybrid training.

To assess the applicability of the framework across modalities, we adopt three domain-appropriate classical backbones and construct their hybrid counterparts by keeping the classical backbone architecture and hyperparameters unchanged, while fusing a projected QCBM embedding into the feature representation immediately before the prediction head. For time-series forecasting, the pure classical model is a two-layer LSTM with hidden size 64, where the last hidden state is fed into a linear head to produce the multivariate forecast. The hybrid LSTM-QCBM model keeps the same LSTM and the same linear head, and only injects the projected QCBM embedding into the last hidden state before the linear head. For computer vision, the pure classical model fine-tunes a ResNet-18 initialized with ImageNet~\cite{deng2009imagenet} pre-trained weights, with the final classifier adapted to the target label space. The hybrid ResNet-18-QCBM model retains the same pre-trained ResNet-18 feature extractor and the same classifier, and fuses the projected QCBM embedding into the feature vector right before the classifier layer. For natural language processing, the pure classical model fine-tunes a pre-trained DistilBERT model for both RTE and TREC with a standard sequence classification head. The hybrid DistilBERT-QCBM model preserves the same DistilBERT backbone and fine-tuning setup, and incorporates the projected QCBM embedding into the pooled sentence representation immediately before the final linear classification head, ensuring that any performance difference is attributable to the injected quantum prior rather than changes in the classical architecture.

\subsubsection{Implementation Details}
All experiments were conducted using PyTorch~\cite{paszke2019pytorch}. The QCBM module was simulated using PennyLane~\cite{bergholm2018pennylane} with a fixed budget of $20{,}480$ measurement shots per circuit evaluation, which are used to estimate the Born distribution $P_{\boldsymbol{\theta}}$ during QCBM training. The use of finite measurement shots allows the simulation to reflect the measurement statistics expected under realistic NISQ deployment. The QCBM parameters were optimized by the gradient-free COBYLA algorithm~\cite{powell1994direct}, with at most $(N_{\theta}+2)$ objective-function evaluations per QCBM update step, where $N_{\theta}$ denotes the number of trainable parameters in the QCBM circuit. For each task, we varied the number of QCBM qubits from 4 to 16 during training and selected the configuration that achieved the best validation performance; the selected qubit number was then fixed for reporting the final test results. For all tasks, the QCBM used a circuit depth of $D=3$, and the task-specific QCBM hyperparameters, including the number of qubits and the fitness score function, are summarized in Table~\ref{tab:QCBM_hyperparams}. When training the Drop-Prior hybrid model, the Bernoulli activation probability of the prior branch was fixed at $p=0.5$ in all experiments, the value at which the Drop-Prior mask variance $p(1-p)$ analyzed in Section~\ref{sec:Drop-Prior Mechanism} is maximized. In time-series forecasting, the models were trained for 10 epochs using Adam~\cite{kingma2015adam} with batch size 32, where the Exchange dataset used an input length of 96 and a prediction length of 96, while the ILI dataset used an input length of 36 and a prediction length of 24. In computer vision, the models were fine-tuned for 20 epochs using Adam with batch size 32. In natural language processing, the DistilBERT model was fine-tuned with batch size 64 for 5 epochs on RTE and 10 epochs on TREC.

\begin{table}[htbp!]
    \centering
    \caption{QCBM hyperparameter settings across datasets. SE denotes Squared Error, and CE denotes Cross-Entropy.}
    \label{tab:QCBM_hyperparams}
    \resizebox{0.7\textwidth}{!}{%
    \begin{tabular}{@{}lccc@{}}
        \toprule
        Dataset & Number of Qubits & Circuit Layers $D$ & Fitness Score Function \\
        \midrule
        Exchange & 16 & 3 & SE \\
        ILI & 6 & 3 & SE \\
        BUSI & 14 & 3 & CE \\
        Aircraft & 8 & 3 & CE \\
        RTE & 6 & 3 & CE \\
        TREC & 12 & 3 & CE \\
        \bottomrule
    \end{tabular}%
    }
\end{table}
\subsection{Main Results}

\subsubsection{Time Series (TS) Forecasting}
We evaluated the framework on two challenging sequential datasets: Exchange and ILI. Exchange exhibits rapid, irregular fluctuations, and due to inherent properties of economic signals, it does not present an obvious periodic pattern; this aperiodicity makes long-horizon forecasting particularly difficult. Similarly, the ILI dataset consists of epidemiological surveillance signals that are also without obvious periodicity at the modeling granularity considered here, and its dynamics are further affected by abrupt distribution shifts caused by evolving public health conditions, which poses substantial challenges for reliable prediction. The classical backbone was an LSTM network, and performance was quantified using Mean Squared Error (MSE) and Mean Absolute Error (MAE).

\begin{table}[H]
    \centering
    \renewcommand{\bfdefault}{b}
    \caption{Performance on Time Series Datasets (Mean $\pm$ Std over 3 seeds). Best results are in bold. Quantum--Classical denotes inference with the QCBM prior enabled; Classical denotes classical inference with the prior removed, where predictions rely solely on the classical parameters learned during hybrid training.}
    \label{tab:ts}
    \resizebox{\textwidth}{!}{%
    \begin{tabular}{@{}ll MM MM@{}}
        \toprule
        \multirow{2}{*}{Model} & \multirow{2}{*}{Inference Mode}
            & \multicolumn{4}{c}{Exchange} & \multicolumn{4}{c}{ILI} \\
        \cmidrule(lr){3-6}\cmidrule(lr){7-10}
        & & \multicolumn{2}{c}{MSE} & \multicolumn{2}{c}{MAE}
          & \multicolumn{2}{c}{MSE} & \multicolumn{2}{c}{MAE} \\
        \midrule
        LSTM & Classical
            & 0.9931 & 0.1104 & 0.7980 & 0.0505 & 1.5230 & 0.2585 & 0.7832 & 0.0614 \\
        \midrule
        \multirow{2}{*}{LSTM-QCBM} & Quantum--Classical
            & 0.8259 & 0.0304 & 0.7496 & 0.0280
            & \bfseries 1.2358 & \bfseries 0.0850 & \bfseries 0.7083 & \bfseries 0.0227 \\
        & Classical
            & \bfseries 0.8209 & \bfseries 0.0301 & \bfseries 0.7483 & \bfseries 0.0283
            & 1.2610 & 0.0698 & 0.7173 & 0.0138 \\
        \bottomrule
    \end{tabular}
    }
\end{table}

Table~\ref{tab:ts} reports test-set results showing consistent improvements of the Drop-Prior Hybrid model over the pure LSTM baseline on both aperiodic datasets. On Exchange, the hybrid model reduces MSE by up to $17.34\%$ (0.9931 $\rightarrow$ 0.8209) and MAE by up to $6.23\%$ (0.7980 $\rightarrow$ 0.7483); on ILI, it reduces MSE by up to $18.86\%$ (1.5230 $\rightarrow$ 1.2358) and MAE by up to $9.56\%$ (0.7832 $\rightarrow$ 0.7083). The hybrid model is also substantially more robust to seed variation: relative to the baseline, the standard deviations of MSE and MAE are reduced under both inference modes, with relative reductions ranging from $43.96\%$ to $77.52\%$. The small performance gap between the two inference modes suggests that the stochastic Drop-Prior training successfully transfers most of the quantum prior's structural information into the classical backbone, enabling accurate and stable prediction even when the quantum prior is entirely removed at deployment time.

\begin{figure}[H]
    \centering
    \begin{subfigure}[b]{0.49\textwidth}
        \centering
        \includegraphics[width=\textwidth]{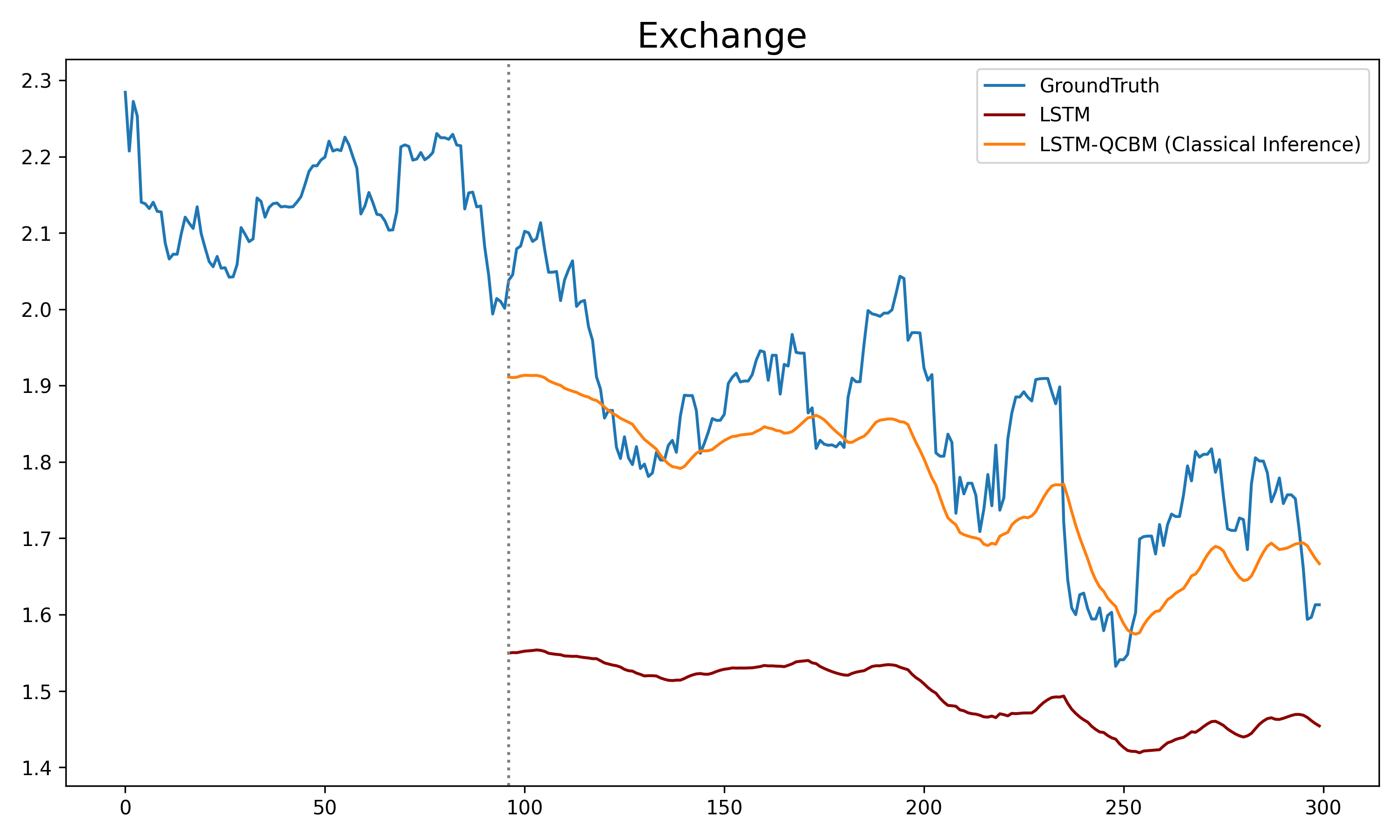}
    \end{subfigure}
    \hfill
    \begin{subfigure}[b]{0.49\textwidth}
        \centering
        \includegraphics[width=\textwidth]{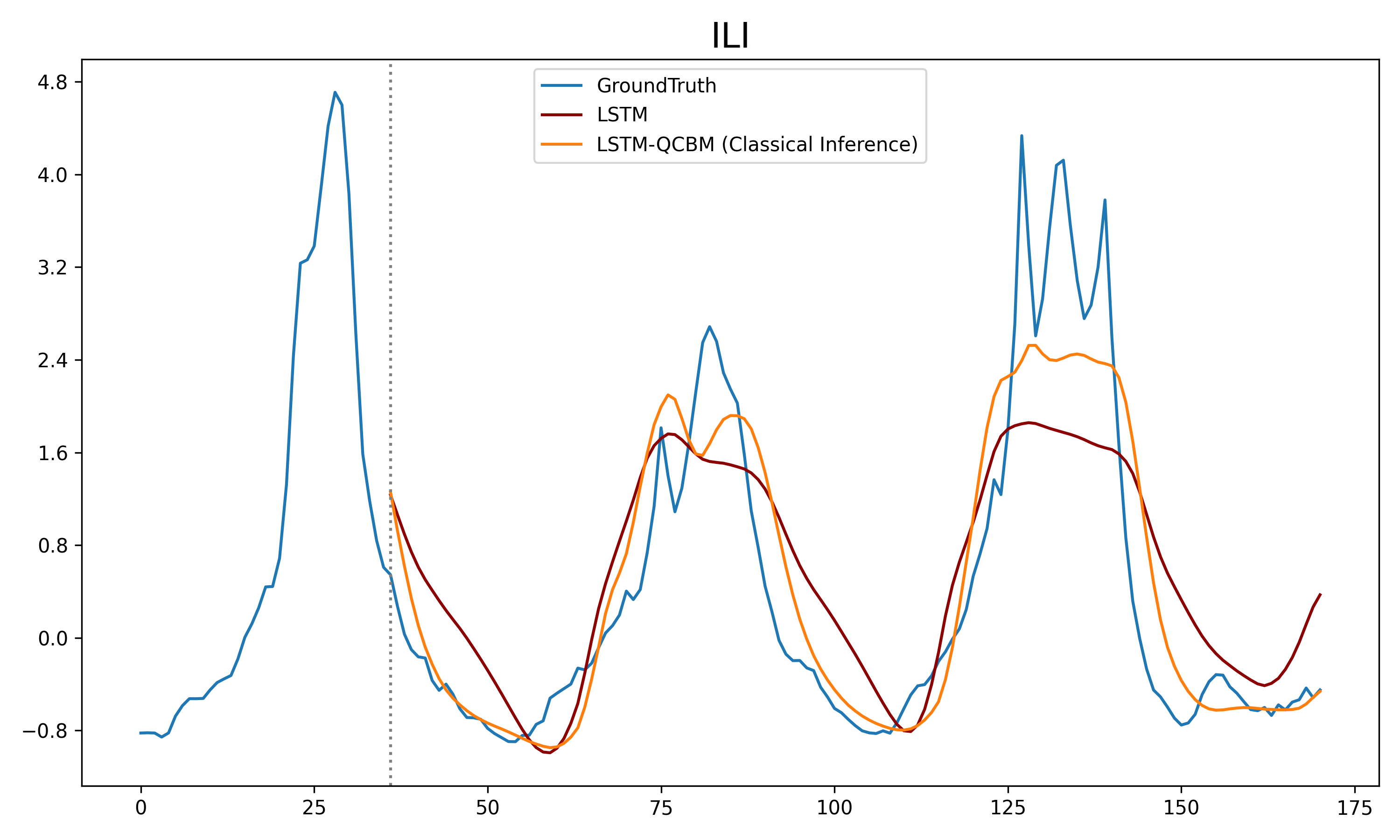}
    \end{subfigure}
    \caption{Visualization of input-96-predict-96 results on the Exchange dataset and input-36-predict-24 results on the ILI dataset. The LSTM-QCBM predictions are produced under \textit{Classical inference}, with the QCBM prior removed.}
    \label{fig:ts_prediction_visualization}
\end{figure}

To further illustrate the improvement in the LSTM's forecasting ability brought by incorporating the QCBM module, Figure~\ref{fig:ts_prediction_visualization} presents the forecasting results on the Exchange and ILI datasets as a qualitative analysis, where LSTM-QCBM discards the quantum prior at test time (\textit{Classical inference}). On Exchange, the classical LSTM produces an overly smooth forecast with a persistent downward bias, whereas LSTM-QCBM more closely follows the ground-truth level, local fluctuations, and major turning points throughout the prediction horizon. On ILI, LSTM-QCBM better captures the timing and overall magnitude of the recurrent peaks and troughs than the LSTM baseline, whose predictions are substantially attenuated and lag the observed dynamics. Although neither model exactly reconstructs abrupt extrema, the QCBM-enhanced model consistently preserves the underlying temporal structure more faithfully. These visual observations complement the quantitative improvements in Table~\ref{tab:ts} and indicate that the learned quantum prior helps the LSTM represent non-stationary, higher-order temporal dependencies while retaining fully classical inference.

\subsubsection{Computer Vision (CV)}
We extended our framework to the CV domain using two image classification datasets: BUSI and Aircraft. The classical backbone was a ResNet-18 architecture initialized from ImageNet pre-trained weights and fine-tuned on each dataset. We evaluated the models based on test Accuracy and Macro-F1 score.

\begin{table}[hptb!]
    \centering
    \renewcommand{\bfdefault}{b}
    \caption{Performance on Computer Vision Datasets (Mean $\pm$ Std over 3 seeds). Best results are in bold. Quantum--Classical denotes inference with the QCBM prior enabled; Classical denotes classical inference with the prior removed, where predictions rely solely on the classical parameters learned during hybrid training.}
    \label{tab:cv}
    \resizebox{\textwidth}{!}{%
    \begin{tabular}{@{}ll MM MM@{}}
        \toprule
        \multirow{2}{*}{Model} & \multirow{2}{*}{Inference Mode}
            & \multicolumn{4}{c}{BUSI} & \multicolumn{4}{c}{Aircraft} \\
        \cmidrule(lr){3-6}\cmidrule(lr){7-10}
        & & \multicolumn{2}{c}{Accuracy} & \multicolumn{2}{c}{Macro-F1}
          & \multicolumn{2}{c}{Accuracy} & \multicolumn{2}{c}{Macro-F1} \\
        \midrule
        ResNet-18 & Classical
            & 0.8456 & 0.0276 & 0.8300 & 0.0300 & 0.3958 & 0.0124 & 0.3889 & 0.0160 \\
        \midrule
        \multirow{2}{*}{ResNet-18-QCBM} & Quantum--Classical
            & 0.8737 & 0.0227 & 0.8561 & 0.0190 & 0.4332 & 0.0045 & 0.4261 & 0.0014 \\
        & Classical
            & \bfseries 0.8772 & \bfseries 0.0248 & \bfseries 0.8613 & \bfseries 0.0184
            & \bfseries 0.4432 & \bfseries 0.0084 & \bfseries 0.4367 & \bfseries 0.0076 \\
        \bottomrule
    \end{tabular}
    }
\end{table}

In computer vision, fine-tuning models on small, domain-specific datasets such as ultrasound images and fine-grained aircraft categories remains susceptible to overfitting even with pre-trained backbones. Classification on the Aircraft subset is particularly challenging because the eight Boeing 737 variants share a highly similar overall appearance and must be distinguished by subtle visual differences. As reported by the test-set results in Table~\ref{tab:cv}, ResNet-18-QCBM consistently outperforms the ResNet-18 baseline on both datasets, reducing the classification error rate by up to $20.47\%$ on BUSI and $7.85\%$ on Aircraft and improving Macro-F1 by up to $3.77\%$ and $12.29\%$ on the two datasets, respectively. ResNet-18-QCBM also exhibits lower variability across the three random seeds: the standard deviations of Accuracy and Macro-F1 fall consistently below those of the baseline under both inference modes, with relative reductions ranging from $10.14\%$ to $91.25\%$ and the most pronounced stability gains observed on Aircraft. The reduced variability is consistent with a regularizing effect of QCBM-assisted training, while the error-rate and Macro-F1 gains indicate that the quantum prior provides a beneficial inductive bias for generalization in these limited-data regimes. The gains retained under \textit{Classical inference} further suggest that useful information acquired during hybrid training is preserved in the ResNet-18 parameters, enabling deployment without the quantum module.

\begin{figure}[H]
    \centering
    \includegraphics[width=\textwidth]{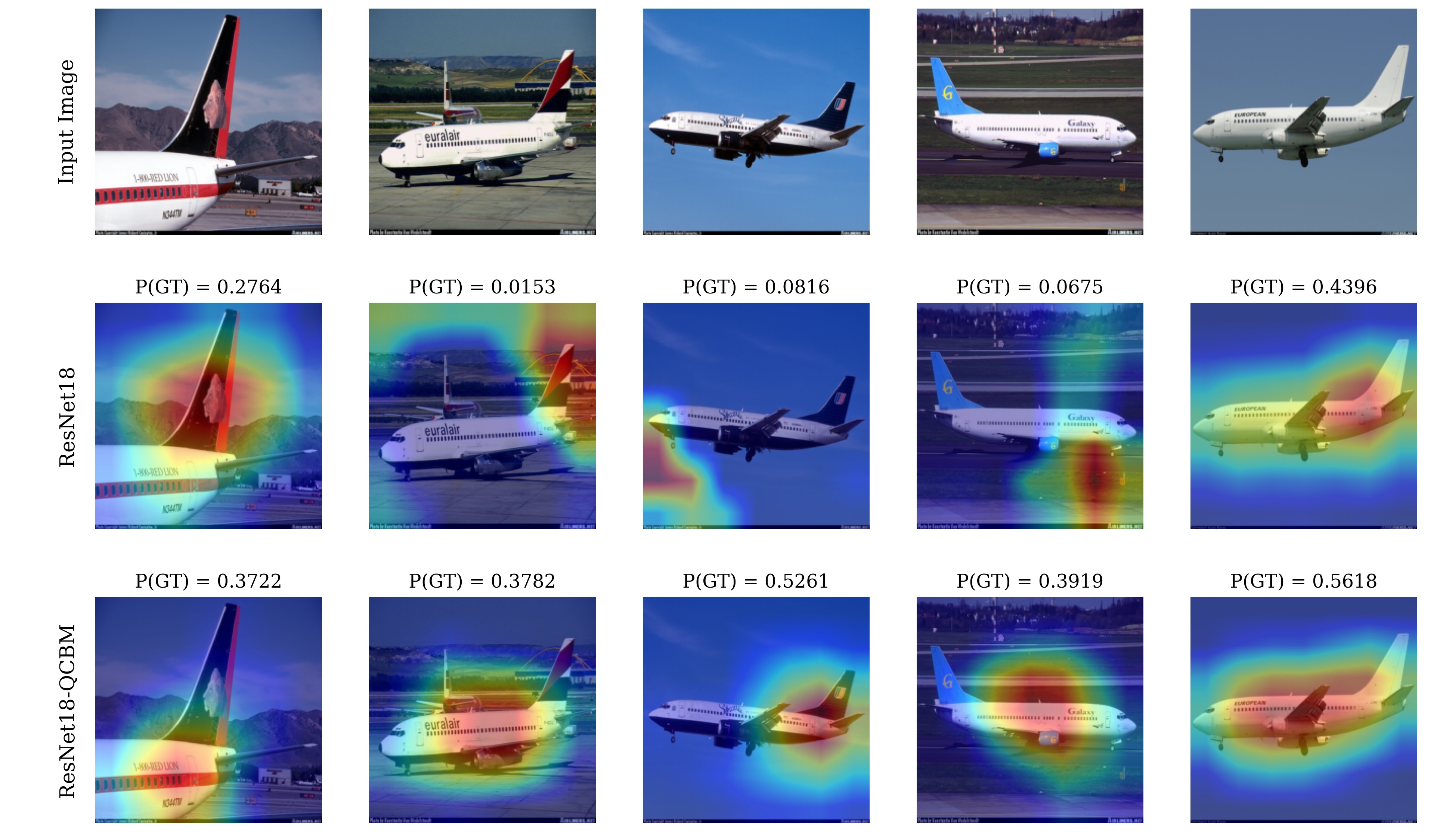}
    \caption{Grad-CAM visualization results on the Aircraft dataset. The first row shows the input images, while the second and third rows show the Grad-CAM maps for ResNet-18 and ResNet-18-QCBM, respectively, where ResNet-18-QCBM is evaluated under \textit{Classical inference} with the QCBM prior removed. The value above each Grad-CAM map indicates the model's predicted probability for the ground-truth class.}
    \label{fig:grad-cam}
\end{figure}

Figure~\ref{fig:grad-cam} compares Grad-CAM maps for five Aircraft images under the ResNet-18 baseline and ResNet-18-QCBM, the latter relying solely on its classical parameters with the quantum prior disabled (\textit{Classical inference}). In several examples, the baseline assigns substantial activation to background regions or only a small part of the aircraft, whereas the QCBM-enhanced model places greater emphasis on aircraft structures such as the fuselage, wings, and tail. The probability assigned to the ground-truth class also increases in all five illustrated cases, most notably from 0.0153 to 0.3782 in the second image and from 0.0816 to 0.5261 in the third. These examples suggest that training with the quantum prior can encourage more discriminative visual representations, helping ResNet-18-QCBM rely more on task-relevant regions when distinguishing fine-grained aircraft variants. The qualitative evidence is consistent with the improved accuracy and Macro-F1 in Table~\ref{tab:cv}.

\subsubsection{Natural Language Processing (NLP)}
For the NLP domain, we evaluated the framework on the RTE and TREC datasets. We utilized a pre-trained DistilBERT model as the classical backbone and constructed its hybrid counterpart, DistilBERT-QCBM, by incorporating the quantum prior module during fine-tuning. Performance was evaluated on the test sets using Accuracy and Macro-F1. Table~\ref{tab:nlp} reports the test-set results for DistilBERT and DistilBERT-QCBM.
\begin{table}[hptb!]
    \centering
    \renewcommand{\bfdefault}{b}   
    \caption{Performance on Natural Language Processing Datasets (Mean $\pm$ Std over 3 seeds). Best results are in bold. Quantum--Classical denotes inference with the QCBM prior enabled; Classical denotes classical inference with the prior removed, where predictions rely solely on the classical parameters learned during hybrid training.}
    \label{tab:nlp}
    \resizebox{\textwidth}{!}{%
    \begin{tabular}{@{}ll MM MM@{}}
        \toprule
        \multirow{2}{*}{Model} & \multirow{2}{*}{Inference Mode}
            & \multicolumn{4}{c}{RTE} & \multicolumn{4}{c}{TREC} \\
        \cmidrule(lr){3-6}\cmidrule(lr){7-10}
        & & \multicolumn{2}{c}{Accuracy} & \multicolumn{2}{c}{Macro-F1}
          & \multicolumn{2}{c}{Accuracy} & \multicolumn{2}{c}{Macro-F1} \\
        \midrule
        DistilBERT & Classical
            & 0.5511 & 0.0385 & 0.5300 & 0.0614 & 0.8933 & 0.0074 & 0.7619 & 0.0290 \\
        \midrule
        \multirow{2}{*}{DistilBERT-QCBM} & Quantum--Classical
            & \bfseries 0.6017 & \bfseries 0.0045 & \bfseries 0.5963 & \bfseries 0.0041 & 0.8933 & 0.0068 & 0.7899 & 0.0174 \\
        & Classical
            &  0.5969 &  0.0017 &  0.5864 &  0.0071
            & \bfseries 0.8960 & \bfseries 0.0075 & \bfseries 0.8060 & \bfseries 0.0069 \\
        \bottomrule
    \end{tabular}%
    }
\end{table}
On RTE, DistilBERT-QCBM improves Accuracy by $9.18\%$ and Macro-F1 by $12.51\%$ under \textit{Quantum--Classical} inference, and retains gains of $8.31\%$ and $10.64\%$ on the two metrics under \textit{Classical} inference relative to the pure DistilBERT baseline. On TREC, Macro-F1 improves by $3.68\%$ and $5.79\%$ under \textit{Quantum--Classical} and \textit{Classical} inference, respectively, demonstrating consistent Macro-F1 gains across both inference modes. DistilBERT-QCBM also reduces the variability across random seeds: on RTE, the standard deviation of Macro-F1 decreases by $88.44\%$--$93.32\%$ relative to the baseline, and on TREC it is reduced by $40.00\%$ and $76.21\%$ under \textit{Quantum--Classical} and \textit{Classical} inference, respectively. Together, the results support the role of the QCBM prior as a task-aligned structured inductive bias that improves downstream generalization. The gains retained after removing the quantum prior are consistent with the intended implicit distillation effect of Drop-Prior training: the classical DistilBERT parameters can preserve useful information acquired during hybrid training without requiring quantum computation at deployment.

\begin{figure}[htbp!]
    \centering
\includegraphics[width=\textwidth]{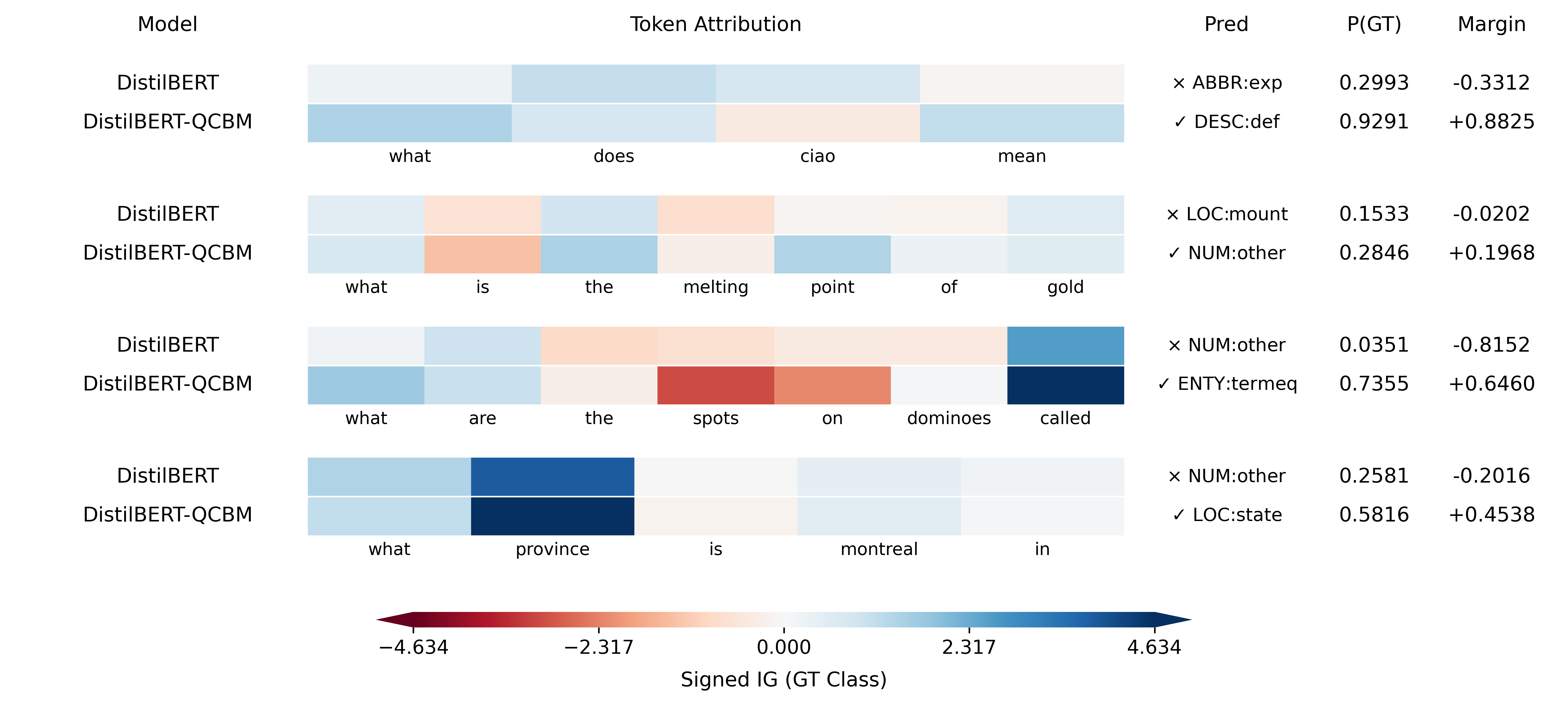}
    \caption{Integrated Gradients (IG) attribution and predictions for four selected error-correction cases from TREC. Each case contains two rows comparing DistilBERT and DistilBERT-QCBM, where DistilBERT-QCBM is evaluated under \textit{Classical inference} with the QCBM prior removed. Left: signed IG attributions to the ground-truth (GT) class logit, with blue and red indicating positive and negative contributions relative to the attribution baseline, respectively. Right: the predicted label (Pred), GT-class probability (P(GT)), and margin, defined as the GT-class probability minus the highest probability among competing classes. A positive margin indicates that the GT class ranks first. Checkmarks ($\checkmark$) and crosses ($\times$) denote correct and incorrect predictions, respectively.}
    \label{fig:ig_attribution}
\end{figure}

To qualitatively explore the token-level evidence associated with the performance gains, Figure~\ref{fig:ig_attribution} presents a case study comparing the Integrated Gradients (IG) attributions and predictions of DistilBERT (baseline) and DistilBERT-QCBM on four examples from the TREC dataset, with DistilBERT-QCBM operating in its prior-free \textit{Classical inference} mode. For each example, we report the ground-truth class probability P(GT) and the prediction margin, defined as P(GT) minus the highest probability among the competing classes, so that a positive margin indicates a correct prediction. The baseline model fails in all four cases, yielding low P(GT) and negative margins. In contrast, DistilBERT-QCBM successfully corrects all misclassifications, significantly boosting the probability of the ground-truth class and achieving highly positive margins. Qualitatively, the attribution heatmaps provide interpretable evidence for this improvement: the baseline model is often distracted by misleading or irrelevant tokens, whereas DistilBERT-QCBM effectively suppresses such noise and assigns strong positive attributions to the genuinely decisive features.
For instance, in the query ``what does ciao mean'' (GT: DESC:def), the baseline overlooks the key intent word ``mean'' and is misled by the token ``ciao'' (predicting ABBR:exp). DistilBERT-QCBM instead captures the ``what...mean'' structural pattern, assigning strong positive attribution to ``mean'' while suppressing ``ciao'' to avoid relying on surface forms. Similarly, for ``what are the spots on dominoes called'' (GT: ENTY:termeq), the baseline is overpowered by the counting semantics of ``spots on dominoes.'' DistilBERT-QCBM, however, explicitly assigns strong negative attributions to these distractors while amplifying the decisive clue ``called'' to the maximum level. As a result, the baseline predicts NUM:other, whereas DistilBERT-QCBM predicts the correct class and increases P(GT) from 0.0351 to 0.7355.
Overall, these cases demonstrate that DistilBERT-QCBM not only substantially improves classification accuracy and confidence but also aligns its decision-making process with more robust and semantically meaningful token evidence.

\subsection{Ablation Study}
\label{sec:ablation}
To assess whether the benefits of the QCBM prior can be reproduced by unstructured randomness or classical generative modeling, we replace it with a uniform prior over bitstrings and a learnable Restricted Boltzmann Machine (RBM)~\cite{fischer2012rbm} prior. We conduct this ablation on one representative dataset per domain, namely ILI for time-series forecasting, BUSI for computer vision, and TREC for natural language processing, so that the comparison spans all three modalities studied in this work while keeping the cost of retraining the fitness-guided prior-learning pipeline for each variant and random seed manageable. The uniform prior samples each bit independently from a Bernoulli distribution with probability $0.5$, assigning equal probability to all strings of a given length. Prior strings are assigned to training samples at initialization and sampled anew for validation and testing, while the distribution remains fixed without fitness-based optimization. In contrast, the RBM represents prior strings through binary visible units, with its hidden-unit count chosen to approximately match the QCBM parameter count at the same string length. It uses exact sampling by enumerating all visible configurations, analytically marginalizing the hidden units, and drawing from the resulting normalized visible-state distribution, with the number of samples drawn per training step fixed at $20{,}480$ to match the QCBM shot budget. Both variants retain the backbone training protocol, prior-fusion architecture, and Drop-Prior mechanism of the QCBM model, while the RBM also follows the same alternating, fitness-guided prior-learning framework. For each variant, prior dimensionality, defined as the binary-string length and corresponding to the QCBM qubit count, is selected independently using validation performance. The selected dimensions for ILI, BUSI, and TREC are 9, 8, and 5 for the uniform prior and 11, 14, and 12 for the RBM prior, respectively. All results are reported under \textit{Classical inference}, with the prior removed at deployment, reflecting the deployment-oriented motivation of the framework.

Table~\ref{tab:ablation} reports test-set results, showing that replacing the QCBM prior with the uniform prior degrades mean performance on all reported metrics across ILI, BUSI, and TREC. Specifically, models trained with the uniform prior exhibit a $31.93\%$ increase in MSE on ILI and a $10.17\%$ decrease in Macro-F1 on TREC relative to their QCBM-prior counterparts. The concurrent decline in Accuracy and Macro-F1 on BUSI further indicates that the effect extends across all three domains. These results suggest that unstructured random prior injection alone cannot reproduce the benefits of QCBM-assisted training. Introducing the learnable RBM prior partially alleviates this deterioration, improving both metrics on ILI and BUSI and Macro-F1 on TREC compared with the uniform prior, although TREC Accuracy does not improve. Nevertheless, the RBM remains inferior to QCBM on all reported mean metrics: MSE on ILI is still $25.97\%$ higher, while Macro-F1 on TREC is $8.60\%$ lower.
\begin{table}[htbp!]
    \centering
    \renewcommand{\bfdefault}{b}
    \caption{Comparison of QCBM, Uniform, and RBM priors on ILI (TS), BUSI (CV), and TREC (NLP) under \textit{Classical inference} (Mean $\pm$ Std over 3 seeds), where the prior is removed at deployment and predictions rely solely on the classical parameters learned during hybrid training. TS, CV, and NLP denote time-series forecasting, computer vision, and natural language processing, respectively. Best results are in bold.}
    \label{tab:ablation}
    \resizebox{\textwidth}{!}{%
    \begin{tabular}{@{}l MM MM MM@{}}
        \toprule
        \multirow{2}{*}{Prior}
            & \multicolumn{4}{c}{TS: ILI}
            & \multicolumn{4}{c}{CV: BUSI}
            & \multicolumn{4}{c}{NLP: TREC} \\
        \cmidrule(lr){2-5}\cmidrule(lr){6-9}\cmidrule(lr){10-13}
        & \multicolumn{2}{c}{MSE} & \multicolumn{2}{c}{MAE}
        & \multicolumn{2}{c}{Accuracy} & \multicolumn{2}{c}{Macro-F1}
        & \multicolumn{2}{c}{Accuracy} & \multicolumn{2}{c}{Macro-F1} \\
        \midrule
        Uniform Prior
            & 1.6637 & 0.1465 & 0.8188 & 0.0424
            & 0.8296 & 0.0179 & 0.8048 & 0.0155
            & 0.8820 & 0.0074 & 0.7240 & 0.0209 \\
        RBM Prior
            & 1.5885 & 0.1539 & 0.7889 & 0.0387
            & 0.8351 & 0.0406 & 0.8155 & 0.0331
            & 0.8807 & 0.0050 & 0.7367 & 0.0224 \\
        \textbf{QCBM Prior}
            & \bfseries 1.2610 & \bfseries 0.0698 & \bfseries 0.7173 & \bfseries 0.0138
            & \bfseries 0.8772 & \bfseries 0.0248 & \bfseries 0.8613 & \bfseries 0.0184
            & \bfseries 0.8960 & \bfseries 0.0075 & \bfseries 0.8060 & \bfseries 0.0069 \\
        \bottomrule
    \end{tabular}%
    }
\end{table}

These findings suggest that the benefits of QCBM-assisted training are
not fully reproduced by random prior injection or the tested classical generative alternative.
Since all three variants share the same backbone training protocol, prior-fusion architecture, and Drop-Prior mechanism, these performance differences originate from the injected prior distribution itself. The uniform prior contains no learnable structure: its samples are drawn independently of the training objective, so the prior branch injects essentially structureless noise during hybrid training and provides no useful information for the Drop-Prior mechanism to distill into the classical weights. The fitness-guided RBM does learn a task-dependent distribution over bitstrings, which accounts for its partial recovery over the uniform prior. Nevertheless, under the same alternating training protocol and with a parameter budget comparable to or smaller than that of the RBM, the QCBM parameterizes a distribution over $2^n$ bitstrings through quantum interference, affording a compact representation of high-order correlations that the tested classical alternative captures less efficiently.
Since the prior is disabled during evaluation, the observed differences reflect benefits retained
in the classical predictor, consistent with the intended implicit distillation effect of Drop-Prior
training.

\section{Conclusion and Discussion}
We have presented QPAT, a universal hybrid quantum-classical framework that successfully overcomes the key barriers of current QML approaches, including the data-loading bottleneck, the quantum--classical interaction and communication overhead during training, and real-time inference latency. The framework introduces three key innovations. First, it treats the QCBM as a learnable generative prior that is fused with classical representations in a modality-agnostic manner, enabling seamless integration into arbitrary classical backbones, including LSTMs, CNNs, and Transformers. Second, a decoupled fitness-based scoring mechanism eliminates gradient coupling between the quantum and classical modules, removing the need for differentiable quantum models and for encoding classical data into quantum states. Third, the stochastic Drop-Prior mechanism implicitly distills the structural knowledge of the quantum prior into the classical weights during training, which our analysis interprets as a noise-driven regularization that flattens the loss landscape and mitigates shortcut learning.

We conducted a series of experiments on six benchmarks spanning TS forecasting, CV, and NLP. QPAT consistently outperforms purely classical backbones, reducing forecasting MSE by up to $18.86\%$ and the classification error rate by up to $20.47\%$, and improving Macro-F1 by up to $12.51\%$, while also surpassing uniform and RBM priors and reducing performance variance across random seeds. The consistent performance ordering over both classical alternatives supports the view that the observed gains stem from the task-aligned structure of the learned Born distribution, rather than from prior injection alone. Crucially, these gains are almost fully retained when the quantum module is entirely removed at inference, confirming that the benefits acquired during hybrid training are preserved in the classical parameters.

Beyond these empirical gains, the asymmetric ``quantum-assisted training, classical inference'' paradigm adopted by QPAT offers system-level advantages for practical deployment. It bypasses the most severe bottlenecks of current quantum hardware. Physical quantum processors remain expensive, environmentally fragile (e.g., requiring cryogenic cooling), and typically accessible only via cloud services; consequently, querying such hardware for real-time inference on end-user devices would introduce unacceptable latency and operational cost. By executing inference entirely on classical hardware, the trained model can be seamlessly deployed on standard servers, laptops, or embedded devices without any quantum infrastructure, thereby removing a primary obstacle to the commercialization of QML.

Several directions remain for future work. First, the current architecture employs an input-agnostic prior that is additively fused with the classical representation immediately before the task head. Although this design is simple and modality-agnostic, the interaction between the quantum prior and the classical representation is restricted in both position and form. Promising extensions include injecting the prior at different or multiple depths of the backbone, and replacing the element-wise addition with richer nonlinear interaction mechanisms, such as cross-attention, gating modulation, or hypernetworks, to enable more expressive quantum-classical feature interaction. Second, our empirical validation is currently restricted to small- and medium-scale backbones, including LSTM, ResNet-18, and DistilBERT. Extending the QCBM prior to large pre-trained models, such as large language models or vision transformers, particularly during parameter-efficient fine-tuning with methods such as low-rank adaptation (LoRA)~\cite{hu2022lora}, would test whether the flat-minimum regularization induced by Drop-Prior training and the anti-overfitting benefits of the quantum prior persist at very large parameter counts. Third, the present training protocol fixes the Drop-Prior probability $p$ and the QCBM update frequency, and how these factors jointly affect performance has not been systematically investigated. Subsequent studies will examine adaptive strategies, including scheduling $p$ and the prior injection strength across training stages and triggering QCBM updates based on the observed benefit or distribution shift. The goal of this direction is to obtain the largest prior-free performance gain with the least prior sampling and optimization, thereby reducing the training cost of practical classical deployment. Finally, all experiments in this work are conducted on quantum simulators. Extending the training phase to physical quantum hardware is a natural next step, in which the interaction between hardware noise and the Drop-Prior masking noise constitutes an interesting open question.

\bibliographystyle{unsrt}
\bibliography{references}

@article{biamonte2017quantum,
  title={Quantum machine learning},
  author={Biamonte, Jacob and Wittek, Peter and Pancotti, Nicola and Rebentrost, Patrick and Wiebe, Nathan and Lloyd, Seth},
  journal={Nature},
  volume={549},
  number={7671},
  pages={195--202},
  year={2017},
  publisher={Nature Publishing Group UK London}
}

@article{schuld2019quantum,
  title={Quantum machine learning in feature Hilbert spaces},
  author={Schuld, Maria and Killoran, Nathan},
  journal={Physical review letters},
  volume={122},
  number={4},
  pages={040504},
  year={2019},
  publisher={APS}
}

@article{chen2026hybrid,
  title={A Hybrid Quantum Circuit Born Machine Framework for Financial Volatility Forecasting: Quantum-Assisted Training and Classical Inference},
  author={Chen, Yixiong},
  journal={arXiv preprint arXiv:2603.09789},
  year={2026}
}

@article{preskill2018quantum,
  title={Quantum {C}omputing in the {NISQ} era and beyond},
  author={Preskill, John},
  journal={Quantum},
  volume={2},
  pages={79},
  year={2018},
  publisher={Verein zur F{\"o}rderung des Open Access Publizierens in den Quantenwissenschaften}
}

@article{mcclean2018barren,
  title={Barren plateaus in quantum neural network training landscapes},
  author={McClean, Jarrod R and Boixo, Sergio and Smelyanskiy, Vadim and Babbush, Ryan and Neven, Hartmut},
  journal={Nature communications},
  volume={9},
  number={1},
  pages={4812},
  year={2018},
  publisher={Nature Publishing Group UK London}
}

@article{cerezo2021cost,
  title={Cost function dependent barren plateaus in shallow parametrized quantum circuits},
  author={Cerezo, Marco and Sone, Akira and Volkoff, Tyler and Cincio, Lukasz and Coles, Patrick J},
  journal={Nature Communications},
  volume={12},
  number={1},
  pages={1791},
  year={2021},
  publisher={Nature Publishing Group UK London}
}

@article{skolik2021layerwise,
  title={Layerwise learning for quantum neural networks},
  author={Skolik, Andrea and McClean, Jarrod R and Mohseni, Masoud and van der Smagt, Patrick and Leib, Martin},
  journal={Quantum Machine Intelligence},
  volume={3},
  number={1},
  pages={5},
  year={2021},
  publisher={Springer}
}

@article{grant2019initialization,
  title={An initialization strategy for addressing barren plateaus in parametrized quantum circuits},
  author={Grant, Edward and Wossnig, Leonard and Ostaszewski, Mateusz and Benedetti, Marcello},
  journal={Quantum},
  volume={3},
  pages={214},
  year={2019},
  publisher={Verein zur F{\"o}rderung des Open Access Publizierens in den Quantenwissenschaften}
}

@article{sweke2020stochastic,
  title={Stochastic gradient descent for hybrid quantum-classical optimization},
  author={Sweke, Ryan and Bergholm, Ville and F{\"o}sel, Thomas and Piroli, Lorenzo and Br{\'a}dler, Kamil and Killoran, Nathan},
  journal={Quantum},
  volume={4},
  pages={314},
  year={2020},
  publisher={Verein zur F{\"o}rderung des Open Access Publizierens in den Quantenwissenschaften}
}

@article{spall1992multivariate,
  title={Multivariate stochastic approximation using a simultaneous perturbation gradient approximation},
  author={Spall, James C},
  journal={IEEE Transactions on Automatic Control},
  volume={37},
  number={3},
  pages={332--341},
  year={1992},
  publisher={IEEE}
}

@article{mari2020transfer,
  title={Transfer learning in hybrid classical-quantum neural networks},
  author={Mari, Andrea and Bromley, Thomas R and Izaac, Josh and Schuld, Maria and Killoran, Nathan},
  journal={Quantum},
  volume={4},
  pages={340},
  year={2020},
  publisher={Verein zur F{\"o}rderung des Open Access Publizierens in den Quantenwissenschaften}
}

@article{perez2020data,
  title={Data re-uploading for a universal quantum classifier},
  author={P{\'e}rez-Salinas, Adri{\'a}n and Cervera-Lierta, Alba and Gil-Fuster, Elies and Latorre, Jos{\'e} Ignacio},
  journal={Quantum},
  volume={4},
  pages={226},
  year={2020},
  publisher={Verein zur F{\"o}rderung des Open Access Publizierens in den Quantenwissenschaften}
}

@article{huang2024shadows,
  title={Shadows of quantum machine learning},
  author={Jerbi, Sofiene and Gyurik, Casper and Marshall, Simon C. and Molteni, Riccardo and Dunjko, Vedran},
  journal={Nature Communications},
  volume={15},
  number={1},
  pages={5676},
  year={2024},
  doi={10.1038/s41467-024-49877-8}
}

@article{havlicek2019supervised,
  title={Supervised Learning with Quantum-Enhanced Feature Spaces},
  author={Havl{\'\i}{\v{c}}ek, Vojt{\v{e}}ch and C{\'o}rcoles, Antonio D and Temme, Kristan and Harrow, Aram W and Kandala, Abhinav and Chow, Jerry M and Gambetta, Jay M},
  journal={Nature},
  volume={567},
  pages={209--212},
  year={2019},
  publisher={Nature Publishing Group UK London}
}

@article{liu2025quantum,
  title={Quantum Relational Knowledge Distillation},
  author={Liu, Chen-Yu and Chen, Kuan-Cheng and Murota, Keisuke and Chen, Samuel Yen-Chi and Rinaldi, Enrico},
  journal={arXiv preprint arXiv:2508.13054},
  year={2025},
  url={https://arxiv.org/abs/2508.13054}
}

@inproceedings{wang2023explicit,
  title={Explicit and Implicit Knowledge Distillation via Unlabeled Data},
  author={Wang, Yuzheng and Ge, Zuhao and Chen, Zhaoyu and Liu, Xiangjian and Ma, Chuang and Sun, Yunquan and Qi, Lizhe},
  booktitle={ICASSP 2023 - 2023 IEEE International Conference on Acoustics, Speech and Signal Processing (ICASSP)},
  pages={1--5},
  year={2023},
  organization={IEEE}
}

@inproceedings{xue2024icd,
  title={iCD: An Implicit Clustering Distillation Method for Structural Information Mining (Student Abstract)},
  author={Xue, Xiang and Ji, Yatu and Ren, Qing-Dao-Er-Ji and Shi, Bao and Lu, Min and Wu, Nier and Zhuang, Xufei and Xu, Haiteng and Cha, Gan-Qi-Qi-Ge},
  booktitle={Proceedings of the AAAI Conference on Artificial Intelligence},
  year={2024}
}

@article{du2024training,
  title={Training Classical Neural Networks by Quantum Machine Learning},
  author={Liu, Chen-Yu and Kuo, En-Jui and Lin, Chu-Hsuan Abraham and Chen, Sean and Young, Jason Gemsun and Chang, Yeong-Jar and Hsieh, Min-Hsiu},
  journal={arXiv preprint arXiv:2402.16465},
  year={2024},
  url={https://arxiv.org/abs/2402.16465}
}

@article{benedetti2019parameterized,
  title={Parameterized quantum circuits as machine learning models},
  author={Benedetti, Marcello and Lloyd, Erika and Sack, Stefan and Fiorentini, Mattia},
  journal={Quantum Science and Technology},
  volume={4},
  number={4},
  pages={043001},
  year={2019},
  doi={10.1088/2058-9565/ab4eb5}
}

@article{cerezo2022challenges,
  title={Challenges and opportunities in quantum machine learning},
  author={Cerezo, M. and Verdon, Guillaume and Huang, Hsin-Yuan and Cincio, Lukasz and Coles, Patrick J.},
  journal={Nature Computational Science},
  volume={2},
  pages={567--576},
  year={2022},
  doi={10.1038/s43588-022-00311-3}
}

@article{schuld2019evaluating,
  title={Evaluating analytic gradients on quantum hardware},
  author={Schuld, Maria and Bergholm, Ville and Gogolin, Christian and Izaac, Josh and Killoran, Nathan},
  journal={Physical Review A},
  volume={99},
  number={3},
  pages={032331},
  year={2019},
  doi={10.1103/PhysRevA.99.032331}
}

@article{wang2021noise,
  title={Noise-induced barren plateaus in variational quantum algorithms},
  author={Wang, Samson and Fontana, Enrico and Cerezo, M. and Sharma, Kunal and Sone, Akira and Cincio, Lukasz and Coles, Patrick J.},
  journal={Nature Communications},
  volume={12},
  number={1},
  pages={6961},
  year={2021},
  doi={10.1038/s41467-021-27045-6}
}

@article{lecun2015deep,
  title={Deep learning},
  author={LeCun, Yann and Bengio, Yoshua and Hinton, Geoffrey},
  journal={Nature},
  volume={521},
  pages={436--444},
  year={2015},
  doi={10.1038/nature14539}
}

@article{geirhos2020shortcut,
  title={Shortcut learning in deep neural networks},
  author={Geirhos, Robert and Jacobsen, J{\"o}rn-Henrik and Michaelis, Claudio and Zemel, Richard and Brendel, Wieland and Bethge, Matthias and Wichmann, Felix A.},
  journal={Nature Machine Intelligence},
  volume={2},
  number={11},
  pages={665--673},
  year={2020},
  doi={10.1038/s42256-020-00257-z}
}

@article{liu2018differentiable,
  title={Differentiable learning of quantum circuit {Born} machines},
  author={Liu, Jin-Guo and Wang, Lei},
  journal={Physical Review A},
  volume={98},
  number={6},
  pages={062324},
  year={2018},
  doi={10.1103/PhysRevA.98.062324}
}

@article{benedetti2019generative,
  title={A generative modeling approach for benchmarking and training shallow quantum circuits},
  author={Benedetti, Marcello and Garcia-Pintos, Delfina and Perdomo, Oscar and Leyton-Ortega, Vicente and Nam, Yunseong and Perdomo-Ortiz, Alejandro},
  journal={npj Quantum Information},
  volume={5},
  pages={45},
  year={2019},
  doi={10.1038/s41534-019-0157-8}
}

@article{sim2019expressibility,
  title={Expressibility and entangling capability of parameterized quantum circuits for hybrid quantum-classical algorithms},
  author={Sim, Sukin and Johnson, Peter D. and Aspuru-Guzik, Alan},
  journal={Advanced Quantum Technologies},
  volume={2},
  pages={1900070},
  year={2019},
  doi={10.1002/qute.201900070}
}

@article{huang2021power,
  title={Power of data in quantum machine learning},
  author={Huang, Hsin-Yuan and Broughton, Michael and Mohseni, Masoud and Babbush, Ryan and Boixo, Sergio and Neven, Hartmut and McClean, Jarrod R.},
  journal={Nature Communications},
  volume={12},
  pages={2631},
  year={2021},
  doi={10.1038/s41467-021-22539-9}
}

@article{schuld2021encoding,
  title={Effect of data encoding on the expressive power of variational quantum-machine-learning models},
  author={Schuld, Maria and Sweke, Ryan and Meyer, Johannes Jakob},
  journal={Physical Review A},
  volume={103},
  number={3},
  pages={032430},
  year={2021},
  doi={10.1103/PhysRevA.103.032430}
}

@article{schreiber2023surrogates,
  title={Classical surrogates for quantum learning models},
  author={Schreiber, Franz J. and Eisert, Jens and Meyer, Johannes Jakob},
  journal={Physical Review Letters},
  volume={131},
  number={10},
  pages={100803},
  year={2023},
  doi={10.1103/PhysRevLett.131.100803}
}

@article{srivastava2014dropout,
  title={Dropout: A simple way to prevent neural networks from overfitting},
  author={Srivastava, Nitish and Hinton, Geoffrey and Krizhevsky, Alex and Sutskever, Ilya and Salakhutdinov, Ruslan},
  journal={Journal of Machine Learning Research},
  volume={15},
  number={56},
  pages={1929--1958},
  year={2014},
  url={https://jmlr.org/papers/v15/srivastava14a.html}
}

@inproceedings{huang2016stochastic,
  title={Deep networks with stochastic depth},
  author={Huang, Gao and Sun, Yu and Liu, Zhuang and Sedra, Daniel and Weinberger, Kilian Q.},
  booktitle={Computer Vision -- ECCV 2016},
  pages={646--661},
  year={2016},
  publisher={Springer},
  doi={10.1007/978-3-319-46493-0_39}
}

@article{bishop1995noise,
  title={Training with noise is equivalent to {Tikhonov} regularization},
  author={Bishop, Christopher M.},
  journal={Neural Computation},
  volume={7},
  number={1},
  pages={108--116},
  year={1995},
  doi={10.1162/neco.1995.7.1.108}
}

@inproceedings{wager2013dropout,
  title={Dropout training as adaptive regularization},
  author={Wager, Stefan and Wang, Sida and Liang, Percy},
  booktitle={Advances in Neural Information Processing Systems},
  volume={26},
  year={2013},
  url={https://arxiv.org/abs/1307.1493}
}

@article{hinton2015distilling,
  title={Distilling the knowledge in a neural network},
  author={Hinton, Geoffrey and Vinyals, Oriol and Dean, Jeff},
  journal={arXiv preprint arXiv:1503.02531},
  year={2015},
  url={https://arxiv.org/abs/1503.02531}
}

@article{gou2021distillation,
  title={Knowledge distillation: A survey},
  author={Gou, Jianping and Yu, Baosheng and Maybank, Stephen John and Tao, Dacheng},
  journal={International Journal of Computer Vision},
  volume={129},
  pages={1789--1819},
  year={2021},
  doi={10.1007/s11263-021-01453-z}
}

@article{hochreiter1997lstm,
  title={Long short-term memory},
  author={Hochreiter, Sepp and Schmidhuber, J{\"u}rgen},
  journal={Neural Computation},
  volume={9},
  number={8},
  pages={1735--1780},
  year={1997},
  doi={10.1162/neco.1997.9.8.1735}
}

@inproceedings{he2016resnet,
  title={Deep residual learning for image recognition},
  author={He, Kaiming and Zhang, Xiangyu and Ren, Shaoqing and Sun, Jian},
  booktitle={Proceedings of the IEEE Conference on Computer Vision and Pattern Recognition},
  pages={770--778},
  year={2016},
  url={https://openaccess.thecvf.com/content_cvpr_2016/html/He_Deep_Residual_Learning_CVPR_2016_paper.html}
}

@article{sanh2019distilbert,
  title={{DistilBERT}, a distilled version of {BERT}: smaller, faster, cheaper and lighter},
  author={Sanh, Victor and Debut, Lysandre and Chaumond, Julien and Wolf, Thomas},
  journal={arXiv preprint arXiv:1910.01108},
  year={2019},
  url={https://arxiv.org/abs/1910.01108}
}

@inproceedings{hu2022lora,
  title={{LoRA}: Low-Rank Adaptation of Large Language Models},
  author={Hu, Edward J. and Shen, Yelong and Wallis, Phillip and Allen-Zhu, Zong and Li, Yuanzhi and Wang, Shean and Wang, Lu and Chen, Weizhu},
  booktitle={International Conference on Learning Representations},
  year={2022},
  url={https://arxiv.org/abs/2106.09685}
}

@inproceedings{selvaraju2017gradcam,
  title={{Grad-CAM}: Visual explanations from deep networks via gradient-based localization},
  author={Selvaraju, Ramprasaath R. and Cogswell, Michael and Das, Abhishek and Vedantam, Ramakrishna and Parikh, Devi and Batra, Dhruv},
  booktitle={Proceedings of the IEEE International Conference on Computer Vision},
  pages={618--626},
  year={2017},
  url={https://openaccess.thecvf.com/content_iccv_2017/html/Selvaraju_Grad-CAM_Visual_Explanations_ICCV_2017_paper.html}
}

@inproceedings{sundararajan2017axiomatic,
  title={Axiomatic attribution for deep networks},
  author={Sundararajan, Mukund and Taly, Ankur and Yan, Qiqi},
  booktitle={Proceedings of the 34th International Conference on Machine Learning},
  volume={70},
  series={Proceedings of Machine Learning Research},
  pages={3319--3328},
  year={2017},
  publisher={PMLR},
  url={https://proceedings.mlr.press/v70/sundararajan17a.html}
}

@inproceedings{kingma2014vae,
  title={Auto-encoding variational {B}ayes},
  author={Kingma, Diederik P. and Welling, Max},
  booktitle={International Conference on Learning Representations},
  year={2014},
  url={https://arxiv.org/abs/1312.6114}
}

@inproceedings{goodfellow2014gan,
  title={Generative adversarial nets},
  author={Goodfellow, Ian and Pouget-Abadie, Jean and Mirza, Mehdi and Xu, Bing and Warde-Farley, David and Ozair, Sherjil and Courville, Aaron and Bengio, Yoshua},
  booktitle={Advances in Neural Information Processing Systems},
  volume={27},
  year={2014}
}

@inproceedings{vaswani2017attention,
  title={Attention is all you need},
  author={Vaswani, Ashish and Shazeer, Noam and Parmar, Niki and Uszkoreit, Jakob and Jones, Llion and Gomez, Aidan N. and Kaiser, {\L}ukasz and Polosukhin, Illia},
  booktitle={Advances in Neural Information Processing Systems},
  volume={30},
  year={2017}
}

@inproceedings{keskar2017sharp,
  title={On large-batch training for deep learning: Generalization gap and sharp minima},
  author={Keskar, Nitish Shirish and Mudigere, Dheevatsa and Nocedal, Jorge and Smelyanskiy, Mikhail and Tang, Ping Tak Peter},
  booktitle={International Conference on Learning Representations},
  year={2017},
  url={https://arxiv.org/abs/1609.04836}
}

@inproceedings{chaudhari2017entropy,
  title={Entropy-{SGD}: Biasing gradient descent into wide valleys},
  author={Chaudhari, Pratik and Choromanska, Anna and Soatto, Stefano and LeCun, Yann and Baldassi, Carlo and Borgs, Christian and Chayes, Jennifer and Sagun, Levent and Zecchina, Riccardo},
  booktitle={International Conference on Learning Representations},
  year={2017},
  url={https://arxiv.org/abs/1611.01838}
}

@inproceedings{lai2018lstnet,
  title={Modeling long- and short-term temporal patterns with deep neural networks},
  author={Lai, Guokun and Chang, Wei-Cheng and Yang, Yiming and Liu, Hanxiao},
  booktitle={Proceedings of the 41st International ACM SIGIR Conference on Research and Development in Information Retrieval},
  pages={95--104},
  year={2018},
  doi={10.1145/3209978.3210006}
}

@article{aldhabyani2020busi,
  title={Dataset of breast ultrasound images},
  author={Al-Dhabyani, Walid and Gomaa, Mohammed and Khaled, Hussien and Fahmy, Aly},
  journal={Data in Brief},
  volume={28},
  pages={104863},
  year={2020},
  doi={10.1016/j.dib.2019.104863}
}

@article{maji2013aircraft,
  title={Fine-grained visual classification of aircraft},
  author={Maji, Subhransu and Rahtu, Esa and Kannala, Juho and Blaschko, Matthew and Vedaldi, Andrea},
  journal={arXiv preprint arXiv:1306.5151},
  year={2013},
  url={https://arxiv.org/abs/1306.5151}
}

@inproceedings{wang2018glue,
  title={{GLUE}: A multi-task benchmark and analysis platform for natural language understanding},
  author={Wang, Alex and Singh, Amanpreet and Michael, Julian and Hill, Felix and Levy, Omer and Bowman, Samuel R.},
  booktitle={International Conference on Learning Representations},
  year={2019},
  url={https://arxiv.org/abs/1804.07461}
}

@inproceedings{li2002trec,
  title={Learning question classifiers},
  author={Li, Xin and Roth, Dan},
  booktitle={Proceedings of the 19th International Conference on Computational Linguistics (COLING)},
  pages={1--7},
  year={2002},
  doi={10.3115/1072228.1072378}
}

@article{bergholm2018pennylane,
  title={{PennyLane}: Automatic differentiation of hybrid quantum-classical computations},
  author={Bergholm, Ville and Izaac, Josh and Schuld, Maria and Gogolin, Christian and Ahmed, Shahnawaz and Ajith, Vishnu and Alam, M. Sohaib and Alonso-Linaje, Guillermo and AkashNarayanan, B. and Asadi, Ali and others},
  journal={arXiv preprint arXiv:1811.04968},
  year={2018},
  url={https://arxiv.org/abs/1811.04968}
}

@inproceedings{paszke2019pytorch,
  title={{PyTorch}: An imperative style, high-performance deep learning library},
  author={Paszke, Adam and Gross, Sam and Massa, Francisco and Lerer, Adam and Bradbury, James and Chanan, Gregory and Killeen, Trevor and Lin, Zeming and Gimelshein, Natalia and Antiga, Luca and others},
  booktitle={Advances in Neural Information Processing Systems},
  volume={32},
  year={2019}
}

@inproceedings{kingma2015adam,
  title={Adam: A method for stochastic optimization},
  author={Kingma, Diederik P. and Ba, Jimmy},
  booktitle={International Conference on Learning Representations},
  year={2015},
  url={https://arxiv.org/abs/1412.6980}
}

@inproceedings{deng2009imagenet,
  title={{ImageNet}: A large-scale hierarchical image database},
  author={Deng, Jia and Dong, Wei and Socher, Richard and Li, Li-Jia and Li, Kai and Fei-Fei, Li},
  booktitle={Proceedings of the IEEE Conference on Computer Vision and Pattern Recognition},
  pages={248--255},
  year={2009},
  doi={10.1109/CVPR.2009.5206848}
}

@inproceedings{fischer2012rbm,
  title={An introduction to restricted {B}oltzmann machines},
  author={Fischer, Asja and Igel, Christian},
  booktitle={Progress in Pattern Recognition, Image Analysis, Computer Vision, and Applications (CIARP)},
  pages={14--36},
  year={2012},
  publisher={Springer},
  doi={10.1007/978-3-642-33275-3_2}
}

@incollection{powell1994direct,
  title={A direct search optimization method that models the objective and constraint functions by linear interpolation},
  author={Powell, Michael J. D.},
  booktitle={Advances in Optimization and Numerical Analysis},
  pages={51--67},
  year={1994},
  publisher={Springer},
  doi={10.1007/978-94-015-8330-5_4}
}

@article{wang2026deep,
  title={Deep time series models: A comprehensive survey and benchmark},
  author={Wang, Yuxuan and Wu, Haixu and Dong, Jiaxiang and Liu, Yong and Wang, Chen and Long, Mingsheng and Wang, Jianmin},
  journal={IEEE Transactions on Pattern Analysis and Machine Intelligence},
  year={2026},
  publisher={IEEE}
}

\end{document}